\documentclass[12pt]{article}
\usepackage{amsmath,amssymb}
\usepackage{graphicx,color}
\numberwithin{equation}{section}

\def\mint{\int_{-\infty}^\infty\!\cdots\!\int_{-\infty}^\infty}

\def\bra#1{\langle #1 |}
\def\ket#1{| #1 \rangle}
\def\ev#1{\langle #1 \rangle}
\def\ip#1#2{\langle #1 | #2 \rangle}

\DeclareMathOperator{\Ai}{Ai}
\def\({\left(}
\def\){\right)}

\DeclareMathOperator{\Tr}{Tr}

\def\sitarel#1#2{\mathrel{\mathop{\kern0pt #1}\limits_{#2}}}

\newcommand{\eqb}{\begin{eqnarray}}
\newcommand{\eqe}{\end{eqnarray}}
\renewcommand{\thefootnote}{\fnsymbol{footnote}}
\newcounter{aff}
\renewcommand{\theaff}{\fnsymbol{aff}}
\newcommand{\affiliation}[1]{
\setcounter{aff}{#1} $\rule{0em}{1.2ex}^\theaff\hspace{-.4em}$}

\begin{document}
\baselineskip=0.7cm
\begin{titlepage}
\hfill\hfill
\begin{minipage}{1.2in}
TIT/HEP-618  
\end{minipage}

\vspace{0.7cm}
\begin{center}
{\LARGE \bf
Exact Results on the ABJM Fermi Gas
}
\lineskip .6em
\vskip1.2cm
{\large Yasuyuki Hatsuda\footnote[1]{\tt yhatsuda@th.phys.titech.ac.jp},
Sanefumi Moriyama\footnote[2]{\tt moriyama@math.nagoya-u.ac.jp} and
Kazumi Okuyama\footnote[3]{\tt kazumi@azusa.shinshu-u.ac.jp}
}
\vskip 2em
\affiliation{1}
{\normalsize\it Department of Physics, Tokyo Institute of Technology\\
Tokyo 152-8551, Japan} \vskip 1 em
\affiliation{2} {\normalsize\it Kobayashi Maskawa Institute 
\& Graduate School of Mathematics, Nagoya University\\
Nagoya 464-8602, Japan} \vskip 1 em
\affiliation{3} {\normalsize\it Department of Physics, Shinshu University\\
Matsumoto 390-8621, Japan} \vskip 1 em
\vskip 1.5em
\end{center}

\begin{abstract}
\vskip 2ex
\baselineskip=3.5ex
We study the Fermi gas quantum mechanics associated to the ABJM matrix
model.
We develop the method to compute the grand partition function of the
ABJM theory, and compute exactly the partition function $Z(N)$ up to
$N=9$ when the Chern-Simons level $k=1$.
We find that the eigenvalue problem  of this quantum mechanical system
is reduced to the diagonalization of a certain Hankel matrix.
In reducing the number of integrations by commuting coordinates and
momenta, we find an exact relation concerning the grand partition
function, which is interesting on its own right and very helpful for
determining the partition function.
We also study the TBA-type integral equations that allow us to compute
the grand partition function numerically.
Surprisingly, all of our exact results of the partition functions are
written in terms of polynomials of $\pi^{-1}$ with rational
coefficients.
\end{abstract}
\vspace*{\fill}
\noindent
July  2012
\end{titlepage}
\renewcommand{\thefootnote}{\arabic{footnote}}
\setcounter{footnote}{0}
\setcounter{section}{0}

\baselineskip = 3.5ex
\section{Introduction}\label{sec:intro}

M-theory is expected to unify all of our understandings on the
nonperturbative effects of string theory.
Despite its importance, many aspects on this interesting theory await
to be understood.
One of the most mysterious facts on M-theory is the degrees of freedom
of the branes.
From the AdS/CFT correspondence, it is known \cite{KT} that $N$
coincident M2-branes (or M5-branes) have the degrees of freedom of
order $N^{3/2}$ (or $N^3$, respectively).
In the case of D-branes in string theory, the worldvolume theory can be 
described by the super Yang-Mills theory,
which has the degrees of freedom $N^2$.
Compared with such a case, the field theoretical realization on the
branes in M-theory is still unclear.

Recently, there is a breakthrough \cite{ABJM} (along with its
extension with fractional branes \cite{ABJ}) in understanding branes
in M-theory, where it was proposed that the worldvolume theory of
M2-branes is described by the supersymmetric extension of the
Chern-Simons theory.
Concretely speaking, the proposal states that $N$ coincident M2-branes
on the geometry ${\mathbb R}^8/{\mathbb Z}_k$ is described by the
${\mathcal N}=6$ Chern-Simons theory with gauge groups
$U(N)_k\times U(N)_{-k}$ and bifundamental matters $A_{1,2},B_{1,2}$ 
forming the superpotential
$W=\frac{2\pi}{k}\Tr(A_1B_1A_2B_2-A_1B_2A_2B_1)$
where the coupling constant is given by $g_s=\frac{2\pi i}{k}$.
The geometry ${\mathbb R}^8/{\mathbb Z}_k$ in a special case with
$k=1$ is nothing but the flat space, hence it is natural to expect
that the supersymmetry is enhanced to ${\mathcal N}=8$ in that case.
It was then proposed that assuming the existence of the monopole
operator we can construct the ${\mathcal N}=8$ supersymmetry algebra
when $k=1,2$ \cite{BKK,GR,KOS,BK}.

Subsequently, it was shown in \cite{KWYexact} that the localization
technique \cite{P}, which reduces the partition function defined by
infinite-dimensional path integral to a finite-dimensional matrix
model, is applicable to supersymmetric Chern-Simons theories.
This matrix model has the measure of the Chern-Simons type and many
techniques from the Chern-Simons matrix model such as \cite{HY} was
utilized \cite{DMP1} to obtain not only the $N^{3/2}$ behavior of the
free energy in the planar limit but also the shift of 't Hooft
coupling $\frac{N}{k}\to \frac{N}{k}-\frac{1}{24}$ as expected from
the Euler coupling in the gravity dual \cite{BH}.

Besides, it was also noticed that this matrix model is closely related to
the topological string theory on a local Calabi-Yau over the 
Hirzebruch surface
${\mathbb F}_0={\mathbb P}^1\times{\mathbb P}^1$.
Using this relation, we can study the higher genus contribution by
borrowing a celebrated recursive relation, the holomorphic anomaly
equation \cite{BCOV}, from the topological string theory.
Following interesting observations by \cite{DMP2}, it was found in
\cite{FHM} that we obtain the Airy function after summing up the
contributions from all genus.

This result on the Airy function suggests some interesting relations
to previous understanding on M-theory.
In \cite{OVV} it was noticed that the Airy function also appears
in other context of M-theory as the wave-function of the universe.
Further, the integral expression of the Airy function
${\rm Ai}\,(z)=\int_C dt\exp(-zt+\frac{t^3}{3})$, which is a contour
integral of an exponentiated cubic term, looks similar to the partition
function $Z=\int DA\exp(-S[A])$ of the cubic Chern-Simons action
$S[A]$.
Therefore, we are led to expect a deep relation among M-theory,
the Chern-Simons theory and the Airy function.

Another expectation from this result is its implication in
understanding some of hidden structures in the membrane theory.
String theory originates from the Veneziano amplitude in the dual
resonance model, out of which the string worldsheet conformal symmetry
was realized.
Similarly, one might expect to find some hidden symmetry out of this
Airy function and realize some structure for M-theory.
An interesting hidden structure of this Airy function was beautifully
realized by \cite{MPfree}.
It was shown that the partition function of M2-branes (as well as its
generalization with ${\mathcal N}=3$ supersymmetry) can be rewritten
exactly into the partition function of a Fermi gas system with the
density matrix (related to Hamiltonian by $\rho=e^{-H}$)
\begin{equation}
\rho=\frac{1}{\sqrt{2\cosh \frac{q}{2}}}
\frac{1}{2\cosh\frac{p}{2}}\frac{1}{\sqrt{2\cosh \frac{q}{2}}},
\end{equation}
and, with the standard analysis of the statistical mechanics, we can
reproduce the behavior of $N^{3/2}$ and Airy function easily.
Very recently, it was further proposed in \cite{MPinteracting} that
we can generalize this formalism to the ${\mathcal N}=2$ case by
introducing interactions to the Fermi particles and in \cite{KMSS} that
we can extend to the cases with Wilson loops.

This simple rederivation of the Airy function implies the importance
of this Fermi gas formulation.
Since we have already a good control on the perturbative terms using
the statistical mechanical method, as a next step, we would like to
proceed to the study of this quantum mechanics extensively beyond the
perturbation theory and to obtain some non-perturbative information
out of it.

Aside from its possible important message on M-theory, the
mathematical structure in this maximally supersymmetric theory seems
very interesting.
The supergravity analysis \cite{LLM} suggests the BPS condition is
expressed by a continuous Toda equation and the Hamiltonian of this
Fermi gas system takes the form similar to the relativistic Toda
theory \cite{RS,OS}.
Therefore, it is natural to expect some kind of the exactly solvable
structure in this theory.
Some of the exact results was found in \cite{O}.

We would like to make a first step towards the goal to compute the
exact partition function of the ABJM theory for arbitrary $N$ and
arbitrary Chern-Simons level $k$, and to solve the eigenvalue equation
of the corresponding Fermi gas system.
Heading for this goal, we have found some interesting exact results
on this theory as we will report in this paper.

As we will see in the next section, the (grand) partition function of
this Fermi gas system is computed by the trace of the power of the
density matrix.
The main difficulty resides in the computation of convoluting the
matrix element successively,
\begin{align}
\rho^n(q_0,q_n)
=\int \frac{dq_{1}}{2\pi}\frac{dq_{2}}{2\pi}\cdots\frac{dq_{n-1}}{2\pi}
\rho(q_0,q_{1})\rho(q_{1},q_{2})\cdots\rho(q_{n-1},q_n).
\label{eq:rho^n-0}
\end{align}
Following the previous work on the thermodynamic Bethe ansatz (TBA)
equations \cite{TW,Z}, here we are able to express the matrix elements
in terms of more basic vector quantities and reducing the computation
of convoluting matrices into that of convoluting vectors. 

In addition, we have also attempted to reduce the integration by
commuting the coordinate $q$ and the momentum $p$.
We have found that, although the naive commutation relation is
anomalous, interestingly the additional term to the naive commutation
relation can be expressed in terms of a projection operator
$|p=0\rangle\langle q=0|$.
Using this relation, we can express $\Tr\rho^n$ for odd $n$ by some
combinations of
$\langle q=0|(2\cosh q/2)^{-1/2}\rho^m (2\cosh q/2)^{1/2}|p=0\rangle$.
For even $n$, we have found an interesting novel relation which
separates the density matrix into two parities.

We will mostly restrict ourselves to the maximally supersymmetric case
$k=1$.
Most of our following analysis goes straightforwardly to other values
of $k$ though a key relation will only look simple for $k=1$.
The case $k=1$ is particularly interesting from the viewpoint of
holography since the ABJM theory at $k=1$ is conjectured to be dual to
the M-theory on $AdS_4\times S^7$ which is in the truly M-theoretic
regime.
The study of the ABJM matrix model in the large $N$, finite $k$ regime
was initiated by \cite{HKPT} and further studied in \cite{MPfree} from
the Fermi gas approach.
However, the approach of \cite{MPfree} is restricted to the
perturbation around $k=0$.
We would like to study the $k=1$ case exactly.

As a result, we have computed the exact partition function $Z(N)$
up to $N=9$.
We shall summarize our results in the following,
\begin{align}
Z(1)&=\frac{1}{4},
&&\hspace{-30mm}Z(2)=\frac{1}{16\pi},\nonumber \\
Z(3)&=\frac{\pi-3}{2^6\pi},
&&\hspace{-30mm}Z(4)=\frac{10-\pi^2}{2^{10}\pi^2},\nonumber \\
Z(5)&=\frac{-9\pi^2+20\pi+26}{2^{12}\pi^2},
&&\hspace{-30mm}Z(6)=\frac{36\pi^3-121\pi^2+78}
{2^{14} \cdot 3^2\pi^3},\nonumber \\
Z(7)&=\frac{-75\pi^3+193\pi^2+174\pi-126}{2^{16}\cdot 3 \pi^3},
&&\hspace{-30mm}Z(8)=\frac{1053\pi^4-2016\pi^3-4148\pi^2+876}
{2^{21}\cdot 3^2 \pi^4},\nonumber \\
Z(9)&=\frac{5517\pi^4-13480\pi^3-15348\pi^2+8880\pi+4140}
{2^{23}\cdot 3^2 \pi^4}.&&
\end{align}
Surprisingly, all of these results are written in terms of polynomials
of $\pi^{-1}$ with rational coefficients.

This paper is organized as follows.
In Section 2, we present our general procedure to compute the grand
partition function.
Applying the lemma of Tracy-Widom \cite{TW}, we first show that the
power of density matrix $\rho^n(q,q')$ is written in terms of a set of
functions of one variable.
We also show that the eigenvalue problem of $\rho$ is reduced to the
diagonalization of a certain Hankel matrix.
In Section 3, we find that the commutation relation of 
$(2\cosh\frac{q}{2})^{-1}$ and $(2\cosh\frac{p}{2})^{-1}$
appearing in $\rho$ contains an anomalous term, which can be expressed
by a insertion of projection operator.
Then we show that such anomalous commutation relation leads to a
nontrivial relation of the grand partition function.
In Section 4, using our general method, we compute exactly the first
few terms of the Taylor expansion of the grand partition function at
$k=1$.
These are nothing but the partition functions $Z(N)$.
Along the way, we find a non-trivial equation relating the even part
and odd part of density matrix.
In Section 5, some numerical results are given.
We consider the TBA-type integral equation proposed in \cite{Z} to
compute $\Tr \rho^n$.
These integral equations are powerful in the numerical analysis.
We also study the numerical solutions of the eigenvalue problem of the
Fermi gas system.
In Section 6, we compare our results with the known perturbative
results of the ABJM matrix model and find a good agreement
with the Airy function even for small $N$.
We study the non-perturbative corrections
to the free energy, and find numerical evidence
that the leading correction scales as $e^{-2\pi\sqrt{2N}}$.
We also consider the expectation value of $N$ in the grand canonical
ensemble.
We find that the numerical computation using TBA-like equations smoothly
interpolates our exact results and the perturbative results.
Finally, we conclude in Section 7 with discussions and future
directions.
Appendices A, B and C contain some details of the computations skipped
in the text.

\section{Grand partition function of ABJM theory}\label{sec:GPF}
The aim of this section is to present a general prescription which is
helpful for computing the grand partition function of the ABJM
theory.
The concrete computation based on the method given here will be
presented in Section \ref{sec:exact}.

Let us start by reviewing the Fermi gas formalism developed in
\cite{MPfree}.
By using the localization, the partition function of the ABJM theory
on $S^3$ is written as the matrix integral
\begin{align}
Z(N)=\frac{1}{N!^2} \int\!
\frac{d^N\mu}{(2\pi)^N}\frac{d^N\nu}{(2\pi)^N}
\frac{\prod_{i<j} [2\sinh(\frac{\mu_i-\mu_j}{2})]^2
[2\sinh(\frac{\nu_i-\nu_j}{2})]^2}
{\prod_{i,j} [2\cosh(\frac{\mu_i-\nu_j}{2})]^2}
\exp\left[\frac{ik}{4\pi}\sum_{i=1}^N(\mu_i^2-\nu_i^2)\right].
\label{eq:Z_ABJM}
\end{align}
Applying the Cauchy identity to the measure factor and performing the
Gaussian integral over $\mu_i,\nu_i$, this partition function is
recast into the ``mirror'' expression
\cite{KWYduality,O,MPfree}
\begin{align}
Z(N)=\frac{1}{2^NN!} \int\!
\frac{d^Nq}{(2\pi k)^N}
\prod_i\frac{1}{2\cosh\frac{q_i}{2}}\prod_{i<j}\tanh^2\frac{q_i-q_j}{2k}~.
\label{eq:Z_mirror}
\end{align}
The key idea of \cite{MPfree} is to rewrite it as the partition
function of an ideal Fermi gas
\begin{align}
Z(N)=\frac{1}{N!}\sum_{\sigma \in S_N}(-1)^{\epsilon(\sigma)}
\int\! \frac{d^Nq}{(2\pi k)^N} \prod_{i=1}^N \rho(q_i,q_{\sigma(i)}),
\label{eq:Z_ABJM2}
\end{align}
where
\begin{align}
\rho(q_1,q_2)=\frac{1}{(2\cosh \frac{q_1}{2})^{1/2}}
\cdot\frac{1}{2 \cosh(\frac{q_1-q_2}{2k})}
\cdot\frac{1}{(2\cosh \frac{q_2}{2})^{1/2}},
\label{eq:rho}
\end{align}
is understood as the density matrix of this Fermi gas system.
Since the sum over permutations in \eqref{eq:Z_ABJM2} can be also
written as a sum over conjugacy classes of the permutation group, it
is convenient to consider the grand partition function
\begin{align}
\Xi(z)=1+\sum_{N=1}^\infty Z(N)z^N,
\label{eq:GPF_def}
\end{align} 
where $z=e^\mu$ is the fugacity and $\mu$ is the chemical potential.
In our analysis, it is important to notice that the grand partition
function has the following form
\begin{align}
\Xi(z)=\exp \left[ - \sum_{n=1}^\infty \frac{(-z)^n}{n} \Tr \rho^n \right],
\label{eq:GPF}
\end{align}
where $\Tr\rho^n$ is defined by
\begin{align}
\Tr \rho^n =\int_{-\infty}^\infty
\frac{dq_1}{2\pi k} \cdots \frac{dq_n}{2\pi k} \,
\rho(q_1,q_2)\rho(q_2,q_3) \cdots \rho(q_n,q_1).
\label{eq:Trrho^n}
\end{align}
It is easy to see that $\Xi(z)$ can be expressed as the Fredholm
determinant of the kernel $\rho$ \footnote{A similar analysis of the
grand partition function of other matrix models has been appeared in
\cite{Kostov:1995xw,KKN,Hoppe:1999xg}.}
\begin{align}
\Xi(z)=\det(1+z\rho).
\end{align}
Once the grand partition function is known, one can obtain the
partition function by performing the contour integral
\begin{align}
Z(N)=\oint \!\frac{dz}{2\pi i} \frac{\Xi(z)}{z^{N+1}}.
\end{align}

For later analysis, it is convenient to divide the density matrix into
two parts.
Since the density matrix $\rho(q_1,q_2)$ is 
parity-preserving: $\rho(-q_1,-q_2)=\rho(q_1,q_2)$,
we decompose it into the parity even and odd parts
\begin{align}
\rho(q_1,q_2)=\rho_+(q_1,q_2)+\rho_-(q_1,q_2),\quad
\rho_\pm(q_1,q_2)=\frac{\rho(q_1,q_2) \pm \rho(q_1,-q_2)}{2}.
\label{eq:rhopm_def}
\end{align}
By definition, $\rho_\pm(q_1,-q_2)=\pm\rho_\pm(q_1,q_2)$, and thus all
the matrix elements of the product $\rho_+\rho_-$ vanish:
\begin{align}
(\rho_+\rho_-)(q_1,q_2)=\int_{-\infty}^\infty \frac{dq}{2\pi}
\rho_+(q_1,q)\rho_-(q,q_2)=0.
\end{align}
This means that $\Tr\rho^n$ splits into two parts,
\begin{align}
\Tr \rho^n=\Tr \rho_+^n+\Tr \rho_-^n. \label{eq:Trrho^n_split}
\end{align}
and the grand partition function is factorized into
\begin{align}
\Xi(z)=\Xi_+(z) \Xi_-(z),
\label{eq:fac_Xi}
\end{align}
where
\begin{align}
\Xi_+(z)=\det(1+z \rho_+),\quad \Xi_-(z)=\det(1+z \rho_-).
\end{align}
As we will see later, such a decomposition of the density matrix plays
an important role in the analysis of the grand partition function.

\subsection{Method to compute the grand partition function}
Now let us consider the computation of the grand partition function.
As can be seen from \eqref{eq:GPF}, we can compute the grand partition
function if we know $\Tr\rho^n$.
To compute the partition function $Z(N)$, in particular, we need
$\Tr\rho^n$ ($1\leq n \leq N$).
However it is technically difficult to directly compute $\Tr\rho^n$
(or the matrix element $\rho^n(q_1,q_2)$) because it is given by the
multi-variable integral as in \eqref{eq:Trrho^n} (or
\eqref{eq:rho^n-0}).
To overcome this difficulty, we here use the technique proposed in
\cite{TW}, which enables us to compute $\rho^n(q_1,q_2)$ from a series
of functions of one variable.
This technique can be applicable to a wide class of the kernels with
the form
\begin{align}
K(q_1,q_2)=\frac{E(q_1)E(q_2)}{M(q_1)+M(q_2)}.
\end{align}
Here we apply it to the kernels $\rho_\pm$.
Below we concentrate our attention to $k=1$ for simplicity.
It is straightforward to generalize the method here to general
$k\geq 2$.

The important point is that two matrices $\rho_\pm(q_1,q_2)$ have the
following forms\footnote{Note that
\begin{align}
\rho(q_1,q_2)=\frac{e^{\frac{1}{2}q_1}}{\sqrt{2\cosh\frac{q_1}{2}}}
\frac{1}{e^{q_1}+e^{q_2}}
\frac{e^{\frac{1}{2}q_2}}{\sqrt{2\cosh\frac{q_2}{2}}}
=\frac{e^{-\frac{1}{2}q_1}}{\sqrt{2\cosh\frac{q_1}{2}}}
\frac{1}{e^{-q_1}+e^{-q_2}}
\frac{e^{-\frac{1}{2}q_2}}{\sqrt{2\cosh\frac{q_2}{2}}}
\notag
\end{align}
itself is also of this class.}
\begin{align}
\rho_\pm(q_1,q_2)=\frac{E_\pm(q_1)E_\pm(q_2)}{\cosh q_1+\cosh q_2},
\label{eq:rhopm}
\end{align}
where
\begin{align}
E_+(q)=\frac{\cosh \frac{q}{2}}{\sqrt{2\cosh \frac{q}{2}}},\quad
E_-(q)=\frac{\sinh \frac{q}{2}}{\sqrt{2\cosh \frac{q}{2}}}.
\end{align}
We can apply Lemma 1 in \cite{TW} to \eqref{eq:rhopm}.
It is shown that (the generating functions of) the
density matrices are given by
\begin{align}
\biggl[\frac{z \rho_\pm}{1-z^2 \rho_\pm^2}\biggr] (q_1,q_2)
=\frac{Q_\pm(q_1)Q_\pm(q_2)-P_\pm(q_1)P_\pm(q_2)}
{\cosh q_1+\cosh q_2}, \label{eq:Rpm}\\
\biggl[\frac{z^2 \rho_\pm^2}{1-z^2 \rho_\pm^2}\biggr] (q_1,q_2)
=\frac{Q_\pm(q_1)P_\pm(q_2)-P_\pm(q_1)Q_\pm(q_2)}
{\cosh q_1-\cosh q_2}, \label{eq:Spm}
\end{align}
where
\begin{align}
P_\pm(q)=\int \! \frac{dq'}{2\pi}
\biggl[\frac{z \rho_\pm}{1-z^2 \rho_\pm^2}\biggr] (q,q')
\sqrt{z}E_{\pm}(q'),\quad
Q_\pm(q)=\int \! \frac{dq'}{2\pi}
\biggl[\frac{1}{1-z^2 \rho_\pm^2}\biggr] (q,q')
\sqrt{z}E_{\pm}(q').
\end{align}
If we expand $P_\pm(q)$ and $Q_\pm(q)$ as
\begin{align}
P_\pm(q)=\sqrt{z} E_\pm(q) \sum_{k=0}^\infty z^{2k+1} \phi_\pm^{2k+1}(q), \quad
Q_\pm(q)=\sqrt{z} E_\pm(q) \sum_{k=0}^\infty z^{2k} \phi_\pm^{2k}(q),
\label{eq:P+Q+}
\end{align}
where
\begin{align}
\phi_\pm^k(q)=\frac{1}{E_\pm(q)}\int \! \frac{dq'}{2\pi}
\rho_\pm^k(q,q') E_\pm(q'),
\label{eq:phipm-def}
\end{align}
then \eqref{eq:Rpm} and \eqref{eq:Spm} indicate that the density
matrices are given by
\begin{align}
\rho_\pm^{2n+1}(q_1,q_2)&=\frac{E_\pm(q_1)E_\pm(q_2)}
{\cosh q_1+\cosh q_2} 
\sum_{k=0}^{2n} (-1)^k \phi_\pm^k(q_1) \phi_\pm^{2n-k}(q_2),
\label{eq:rho_odd}\\
\rho_\pm^{2n}(q_1,q_2)&=\frac{E_\pm(q_1)E_\pm(q_2)}
{\cosh q_1-\cosh q_2} 
\sum_{k=0}^{2n-1} (-1)^k \phi_\pm^k(q_1) \phi_\pm^{2n-1-k}(q_2).
\label{eq:rho_even}
\end{align}
These are the key relations in our analysis later.
These results show that the matrix elements $\rho_\pm^n(q_1,q_2)$ are
determined algebraically by the functions $\phi_\pm^k(q)$
($k=0,\dots,n-1$) of one variable respectively.
Therefore our problem reduces to finding $\phi_\pm^k(q)$, which is
much simpler than the original one of finding $\rho_\pm^n(q_1,q_2)$
themselves.

For the determination of $\phi_\pm^k(q)$, as discussed in Appendix
\ref{app:rec-phi}, we can show that $\phi_\pm^k(q)$ satisfy the
following recurrence relations
\begin{align}
\phi_+^k(q)=c_q \int_{-\infty}^\infty \frac{dq'}{2\pi} 
c_{q-q'} \phi_+^{k-1}(q'),\quad
\phi_-^k(q)=c_q \int_{-\infty}^\infty \frac{dq'}{2\pi} 
c_{q-q'}t_{q'}^2 \phi_-^{k-1}(q'). 
\label{eq:phipm_rec}
\end{align}
Throughout this paper, we often use the notations 
\begin{align}
c_q=\frac{1}{2\cosh \frac{q}{2}},\quad
s_q=\frac{1}{2\sinh \frac{q}{2}},\quad 
t_q=\tanh \frac{q}{2}.
\end{align}
From \eqref{eq:phipm_rec} with the initial condition
$\phi_\pm^0(q)=1$, one can recursively fix $\phi_\pm^k(q)$.
In Section \ref{sec:exact}, we will indeed determine the first few $\phi_\pm^k(q)$
by using these recurrence relations.

Let us summarize our general prescription to compute the grand
partition function.
We first determine $\phi_\pm^k(q)$ by  using \eqref{eq:phipm_rec}.
Then from \eqref{eq:rho_odd} and \eqref{eq:rho_even}, we find the
matrix elements $\rho_\pm^n(q_1,q_2)$.
Computing $\Tr \rho_\pm^n$ from such expressions, we obtain the grand
partition function through \eqref{eq:GPF} and
\eqref{eq:Trrho^n_split}.

\subsection{Eigenvalue problem and Hankel matrix representation}
\label{sec:eigen}
In the previous subsection, we considered $\Tr \rho^n$ to know the grand
partition function.
There is another approach to analyze it.
The grand partition function can be expressed in terms of the
eigenvalues of the density matrix $\rho$,
\begin{align}
\Xi(z)=\det (1+z \rho)=\prod_{n=0}^\infty (1+z \lambda_n),
\end{align}
where $\lambda_n$ is the $n$-th eigenvalue of $\rho$ with decreasing
order $\lambda_0 \geq \lambda_1\geq \cdots$.
Thus if we can solve the eigenvalue problem of $\rho$, we obtain
$\Xi(z)$.
The eigenvalue equation is given by the homogeneous Fredholm integral
equation of the second kind
\begin{align}
\int_{-\infty}^\infty \frac{dq'}{2\pi} \rho(q,q')\Psi(q')=\lambda \Psi(q).
\label{eq:eigen_eq}
\end{align}
It is, of course, hard to solve this equation, but we can rewrite it
as the eigenvalue equation for an infinite dimensional Hankel matrix
$M$ whose matrix elements are given by
\begin{align}
M_{nm}
=\frac{1}{8\pi k}\int_{-\infty}^\infty dq 
\frac{\tanh^{n+m}(\frac{q}{2k})}{\cosh(\frac{q}{2}) \cosh^2(\frac{q}{2k})}
=\frac{1}{4\pi}\int_{-1}^1 dt \frac{t^{n+m}}{T_k(1/\sqrt{1-t^2})} ,
\label{eq:Hankel1}
\end{align}
where $T_k(x)$ is the $k$-th Chebyshev polynomial of the first kind.
This representation is useful in our later analysis.

To rewrite the eigenvalue equation, let us first expand the kernel
\eqref{eq:rho} into
\begin{align}
\rho(q_1,q_2)=\sum_{n=0}^\infty \rho_n(q_1)\rho_n(q_2),
\label{eq:rho_exp}
\end{align}
where
\begin{align}
\rho_n(q)=\frac{\tanh^n(\frac{q}{2k})}
{2\sqrt{k \cosh(\frac{q}{2})}\cosh(\frac{q}{2k})}.
\end{align}
Since $\rho_n(-q)=(-1)^n \rho_n(q)$, the decomposition
\eqref{eq:rhopm_def} indicates that $\rho_\pm$ are respectively
expressed as
\begin{align}
\rho_+(q_1,q_2)=\sum_{n=0}^\infty \rho_{2n}(q_1)\rho_{2n}(q_2),\quad
\rho_-(q_1,q_2)=\sum_{n=0}^\infty \rho_{2n+1}(q_1)\rho_{2n+1}(q_2)
\end{align}
Plugging \eqref{eq:rho_exp} into \eqref{eq:eigen_eq}, we obtain
\begin{align}
\sum_{n=0}^\infty v_n \rho_n(q)=\lambda \Psi(q),
\label{eq:sum_vn_rhon}
\end{align}
where
\begin{align}
v_n=\int_{-\infty}^\infty \frac{dq}{2\pi} \rho_n(q)\Psi(q).
\label{eq:vn1}
\end{align}
Substituting \eqref{eq:sum_vn_rhon} back to \eqref{eq:vn1}, we find
\begin{align}
\sum_{m=0}^\infty M_{nm} v_m=\lambda v_n,
\label{eq:eigen_M}
\end{align}
where the matrix elements of $M$ are given by \eqref{eq:Hankel1}.
The equation \eqref{eq:eigen_M} formally shows that all the eigenvalues of
$\rho$ are also the eigenvalues of the infinite dimensional matrix
$M$.

The matrix elements $M_{nm}$ depend only on $n+m$, and such matrices
are called Hankel matrices.
For the present case, it is obvious that $M_{nm}=0$ if $n+m$ is odd.
Namely, this Hankel matrix $M$ has the following form
\begin{align}
M=\begin{pmatrix}
m_0&0&m_1&0&m_2&0&\cdots\\
0&m_1&0&m_2&0&m_3&\\
m_1&0&m_2&0&m_3&0&\\
0&m_2&0&m_3&0&m_4&\\
m_2&0&m_3&0&m_4&0\\
\vdots&&&&&&\ddots
\end{pmatrix}.
\end{align}
We shall call such a matrix a parity-preserving Hankel matrix.
As can be seen easily, the parity-preserving Hankel matrix can be
decomposed into two blocks of Hankel matrices with opposite parity
\begin{align}
M_+=\begin{pmatrix}
m_0&m_1&m_2&\cdots\\
m_1&m_2&m_3&\\
m_2&m_3&m_4&\\
\vdots&&&\ddots
\end{pmatrix},
\qquad
M_-=\begin{pmatrix}
m_1&m_2&m_3&\cdots\\
m_2&m_3&m_4&\\
m_3&m_4&m_5&\\
\vdots&&&\ddots
\end{pmatrix}.
\end{align}
These two Hankel matrices are related by
\begin{align}
M_-=M_+T_-=T_+M_+,
\label{eq:Mm-Mp}
\end{align}
with the shift matrix $(T_\pm)_{mn}=\delta_{m\pm 1,n}$.
The eigen-spaces of $M$ also decompose into the direct product of the
eigen-spaces of $M_\pm$.

For $k=1$, $M_{nm}$ has the following very simple form ($n+m$: even)
\begin{align}
M_{nm}^{(k=1)}=\frac{C_{\frac{n+m}{2}}}{2^{n+m+3}},
\label{eq:Hankel_k=1}
\end{align}
where $C_n$ is the Catalan number
\begin{align}
C_n=\frac{(2n)!}{(n+1)!n!}~.
\end{align}
For $k=2$, we find ($n+m$: even)
\begin{align}
M_{nm}^{(k=2)}=\frac{1}{8\pi}
\left[-\frac{2}{n+m+1}
+\psi\(\frac{n+m+3}{4}\)-\psi\(\frac{n+m+1}{4}\)\right],
\end{align}
where $\psi(x)=\Gamma'(x)/\Gamma(x)$ is the digamma function.

After seen the simple Hankel matrix representation, before closing
this subsection, now let us express various quantities appearing in
the previous subsection in terms of this Hankel matrix $M$.
Since the matrix $M$ and the density matrix $\rho$ share the same set
of eigenvalues assuming a certain regularity condition, it is obvious
that $\Tr M^n$ is equal to $\Tr \rho^n$ for any $n$, thus the grand
partition function is given by
\begin{align}
\Xi(z)=\det(1+z M).
\end{align}
Repeating the same argument above, one finds that the eigenvalue
problems for $\rho_{\pm}$ are mapped to those for the Hankel matrices
$M_{\pm}$, respectively.
Let $\lambda_{\pm,n}$ ($n=0,1,2,\dots$) be the $n$-th
eigenvalue of the density matrix $\rho_{\pm}$.
Then, we find
\begin{align}
\Xi_+(z)&=\prod_{n=0}^\infty (1+z\lambda_{+,n})=\det(1+z M_+),\nonumber\\
\Xi_-(z)&=\prod_{n=0}^\infty (1+z\lambda_{-,n})=\det(1+z M_-).
\label{eq:Xi_pm}
\end{align}
We observe from the numerical evaluations (see Section
\ref{sec:num_eigen}) that the eigenvalues of $\rho_{\pm}$ satisfy the
inequality
\begin{align}
\lambda_{+,0}>\lambda_{-,0}
>\lambda_{+,1}>\lambda_{-,1}
>\lambda_{+,2}>\lambda_{-,2}>\cdots,
\end{align}
which means the following relations
\begin{align}
\lambda_{+,n}=\lambda_{2n},\quad \lambda_{-,n}=\lambda_{2n+1}.
\label{eq:lambda_pm}
\end{align}
Also, $\phi_+^k(q)$ and its Fourier transformation
$\tilde{\phi}_+^k(p)$
\begin{align}
\tilde{\phi}_+^k(p)=\int\!\frac{dq}{2\pi} e^{-\frac{i}{2\pi} pq}\phi_+^k(q).
\end{align}
can be expressed in terms of $M$ as
\begin{align}
\frac{1}{4}c_q^{-2} \phi_+^k(q)=\sum_{m,n=0}^\infty t_q^{2n} (M_+^k)_{nm},\quad
2c_q^{-1}\tilde{\phi}_+^k(q)=\sum_{n=0}^\infty t_q^{2n} (M_+^{k-1})_{n0}.
\label{eq:phi-Hankel}
\end{align}
where $t_q=\tanh(q/2)$.
The derivation of these representations will be given in Appendix
\ref{app:phi-Hankel}.

\subsection{Representation in Chebyshev basis}
In the previous subsection, we have found an interesting Hankel matrix
representation for the density matrix $\rho(q_1,q_2)$.
It is natural to wonder whether there is an orthonormal basis in which
the density matrix $\rho(q_1,q_2)$ is represented by this Hankel
matrix.
We will show in this subsection that the Hankel matrix $M$ is closely
related to the density matrix in a basis constructed by the Chebyshev
polynomials.

Motivated by the expansion \eqref{eq:rho_exp}, let us introduce the
following function
\begin{align}
\ip{q}{n}=\frac{\sqrt{2}U_n(\tanh (\frac{q}{2}))}
{\cosh^{3/2} (\frac{q}{2})},
\end{align}
where $U_n(x)$ is the Chebyshev polynomial of the second kind and its
appearance originates from the orthonomality of the basis
$\{ \ket{n} \}_{n=0}^\infty$,
\begin{align}
\ip{m}{n}=\int_{-\infty}^\infty \frac{dq}{2\pi}\, 
\frac{\sqrt{2}U_m(\tanh (\frac{q}{2}))}{\cosh^{3/2} (\frac{q}{2})}
\frac{\sqrt{2}U_n(\tanh (\frac{q}{2}))}{\cosh^{3/2} (\frac{q}{2})}
=\delta_{mn}. 
\end{align} 
Then, the density matrix in this basis,
$M_{n_1n_2}^\rho=\bra{n_1}\rho\ket{n_2}$, is given by
\begin{align}
M_{n_1n_2}^\rho
&=\int\!\frac{dq_1}{2\pi}\frac{dq_2}{2\pi}
\ip{n_1}{q_1}\bra{q_1}\rho \ket{q_2} \ip{q_2}{n_2}
=\sum_{n=0}^\infty \int\! \frac{dq_1}{2\pi}\ip{q_1}{n_1} \rho_n(q_1) 
\int\! \frac{dq_2}{2\pi}\ip{q_2}{n_2} \rho_n(q_2). 
\end{align}
where we have used the expansion \eqref{eq:rho_exp}.
Let $B$ be the matrix for the basis change from Chebyshev polynomials
to monomials
\begin{align}
t^n=\sum_{m=0}^\infty B_{nm}U_m(t).
\end{align}
The explicit form of the non-vanishing matrix element $B_{nm}$ is
given by
\begin{align}
B_{nm}=\frac{1}{2^n}\frac{m+1}{n+1} \binom{n+1}{(n-m)/2},
\end{align}
when $n+m$ is even and $n \geq m$.
Since one can check
\begin{align}
\int\! \frac{dq_1}{2\pi}\ip{q_1}{n_1} \rho_n(q_1)
=\sum_m B_{nm}\int\! \frac{dq_1}{2\pi}
\frac{U_{n_1}(\tanh(\frac{q_1}{2}))U_m(\tanh(\frac{q_1}{2}))}
{\sqrt{2}\cosh^3(\frac{q_1}{2})}
=\frac{1}{2\sqrt{2}}B_{nn_1},
\end{align}
the matrix element $M_{n_1n_2}^\rho$ is given by
\begin{align}
M_{n_1n_2}^\rho=\frac{1}{8}\sum_{n=0}^\infty B_{nn_1}B_{nn_2}=\frac{1}{8}(B^T B)_{n_1n_2}.
\end{align}
On the other hand, the Hankel matrix $M$ for $k=1$ is written as
\begin{align}
M_{n_1 n_2}&=\frac{1}{8\pi} \int\! 
dq \frac{\tanh^{n_1+n_2}(\frac{q}{2})}{\cosh^3(\frac{q}{2})}
=\frac{1}{8\pi}\sum_{m_1,m_2}B_{n_1m_1}B_{n_2m_2}\int\! dq 
\frac{U_{m_1}(\tanh(\frac{q}{2}))U_{m_2}(\tanh(\frac{q}{2}))}
{\cosh^3(\frac{q}{2})}\notag \\
&=\frac{1}{8}(BB^T)_{n_1n_2}.
\end{align}
In summary, the Hankel matrix $M$ and the density matrix $M^\rho$ in
the Chebyshev basis are written as
\begin{align}
M=\frac{1}{8}BB^T,\qquad M^\rho=\frac{1}{8}B^T B.
\end{align}
These two matrices are isospectral, {\it i.e.}, they have the same set
of eigenvalues (assuming a certain regularity condition)
\begin{align}
Mv=\lambda v\;\; \Leftrightarrow\;\; M^\rho v' = \lambda v',
\qquad v'=B^T v.
\end{align}
Note that $M^\rho$ is not a Hankel matrix while $M$ is so.

\section{Algebraic treatment}\label{sec:algebraic}
Here we consider an algebraic treatment of our Fermi gas system.
Note that the matrix element in the position representation
\eqref{eq:rho}
\begin{align}
\rho(q_1,q_2)=\bra{q_1} \hat{\rho} \ket{q_2},
\end{align}
comes from a one-dimensional quantum mechanical system with the
operator $\hat{\rho}$ expressed in terms of the coordinate $\hat{q}$
and the momentum $\hat{p}$ as
\begin{align}
\hat{\rho}=\sqrt{c_{\hat{q}}}c_{\hat{p}} \sqrt{c_{\hat{q}}}.
\end{align}
Below we focus on the $k=1$ case.
Let us summarize our convention for the phase space used in this
paper, which is especially clean for $k=1$.
The commutation relation of $\hat{q}$ and $\hat{p}$ is given by
\begin{align}
[\hat{q},\hat{p}]=2\pi i.
\end{align}
The normalization of the bases $\ket{q}$ and $\ket{p}$ is as follows:
\begin{align}
\ip{q}{q'}=2\pi \delta(q-q'),\quad \ip{p}{p'}=2\pi\delta(p-p'),\quad
\ip{q}{p}=e^{\frac{i}{2\pi}pq}.
\end{align}
Therefore the projectors are normalized as
\begin{align}
1=\int_{-\infty}^\infty \frac{dq}{2\pi} \ket{q}\bra{q}
=\int_{-\infty}^\infty \frac{dp}{2\pi} \ket{p}\bra{p}.
\end{align}
Hereafter, we suppress the hats on operators for simplicity.
Note that the matrix elements of $c_p$ and $s_p$ in the position
representation are
\begin{align}
\bra{q_1} c_p \ket{q_2}=c_{q_1-q_2},\quad 
\bra{q_1} s_p \ket{q_2}=\frac{i}{2} \tanh \( \frac{q_1-q_2}{2} \).
\label{eq:cp-sp_ele}
\end{align}
Using the bra-ket notation, 
the matrix elements of $\rho^n$ are given by
\begin{align}
\rho^n(q_1,q_2)=\bra{q_1}(\sqrt{c_q}c_p\sqrt{c_q})^n\ket{q_2}
=\sqrt{\frac{c_{q_1}}{c_{q_2}}}\,\bra{q_1}(c_p c_q)^n\ket{q_2}.
\label{eq:rho^n_ele}
\end{align}
Since $\Tr \rho^n=\Tr [(c_pc_q)^n ]=\Tr [(c_qc_p)^n]$ from this
equation, the grand partition function is also written as
\begin{align}
\Xi(z)=\det(1+z c_p c_q)=\det(1+z c_q c_p). 
\end{align}

\subsection{Anomalous terms in commutation relations}
As explained around \eqref{eq:rho^n-0}, the difficulty in computing
$\Tr\rho^n$ consists in the multi-dimensional integral.
In the operator representation, it can also be stated as the difficulty
in the integration over the multi-dimensional phase space.
Therefore, we would like to reduce the dimension.

If we know the commutation relation between $c_p$ and $c_q$, we can
reduce the dimension of the phase space involved.
As is seen here, one should be careful to derive the commutation
relation.
From the operator identity
\begin{align}
2\cosh \frac{q}{2} \cdot 2\cosh \frac{p}{2}
=i\cdot 2 \sinh \frac{p}{2} \cdot 2 \sinh \frac{q}{2},
\end{align} 
one would naively expect
\begin{align}
\frac{1}{2\cosh \frac{p}{2}} \cdot \frac{1}{2\cosh \frac{q}{2}}
\stackrel{?}{=}-i\cdot
\frac{1}{2\sinh \frac{q}{2}} \cdot \frac{1}{2\sinh \frac{p}{2}}.
\label{eq:naive_com}
\end{align}
However this is not the case.
There is an anomalous term in this commutation relation.
To see this, let us consider the matrix element of 
the difference between both sides of \eqref{eq:naive_com}
\begin{align}
\bra{q_1}\frac{1}{2\cosh \frac{p}{2}}
\cdot \frac{1}{2\cosh \frac{q}{2}}\ket{q_2}
=-i\bra{q_1}\frac{1}{2\sinh \frac{q}{2}}
\cdot \frac{1}{2\sinh \frac{p}{2}}\ket{q_2}
+\frac{1}{4\sinh \frac{q_1}{2}}\tanh\frac{q_2}{2}.
\end{align}
If we introduce a projection operator
$\ket{p=0}\bra{q=0}=:\ket{0_p}\bra{0_q}$ satisfying
$(\ket{0_p}\bra{0_q})^2=\ket{0_p}\bra{0_q}$, we can express the
relation as
\begin{align}
c_p c_q=-i s_q \Pi s_p,\quad \Pi =1-\ket{0_p}\bra{0_q},
\label{eq:cpcq_comm}
\end{align}
at the operator level.
Similarly, we find
\begin{align}
c_q c_p=i s_p \Pi^\dagger s_q,\quad \Pi^\dagger =1-\ket{0_q}\bra{0_p}.
\label{eq:cqcp_comm}
\end{align}

Using the commutation relations \eqref{eq:cpcq_comm} and
\eqref{eq:cqcp_comm}, we can calculate $\Tr \rho^n$ correctly.
Let us see the computation of $\Tr \rho^2$ as a simple example:
\begin{align}
\Tr \rho^2=\Tr (c_p c_q c_p c_q)
=-i \Tr (s_q \Pi s_p c_p c_q)
=-i \Tr (s_{2q} \Pi s_{2p}),
\label{eq:Trrho^2}
\end{align}
where we have used $s_q c_q =s_{2q}$.
Since $\Pi=1-\ket{0_p}\bra{0_q}$, \eqref{eq:Trrho^2} is further
rewritten as
\begin{align}
\Tr \rho^2=-i \int\! \frac{dqdp}{(2\pi)^2} s_{2q}s_{2p}
+i\int\! \frac{dq}{2\pi} 
\bra{q}s_{2q} \ket{0_p}\bra{0_q}s_{2p} \ket{q}.
\end{align}
The first term vanishes because the integrand is an odd function.
The second term is evaluated as
\begin{align}
i\int\! \frac{dq}{2\pi} \bra{q}s_{2q} \ket{0_p}\bra{0_q}s_{2p} \ket{q}
=\frac{1}{4}\int\! \frac{dq}{2\pi} c_q c_{q/2}^2.
\end{align}
where $\bra{q}s_{2q} \ket{0_p}=s_{2q}$ and
$\bra{0_q}s_{2p} \ket{q}=-(i/4) \tanh(q/4)$.
Therefore we obtain
\begin{align}
\Tr \rho^2=\frac{\pi-2}{16\pi}.
\end{align}
One can easily check that the same result is reproduced by a brute
force integration using \eqref{eq:Trrho^n}.
In the above computation, it is obvious that the anomalous term in the
commutation relation is crucial to obtain the correct answer.
In principle, we can compute $\Tr \rho^n$ by using the relations
\eqref{eq:cpcq_comm} and \eqref{eq:cqcp_comm} repeatedly instead of
doing multi-integral.
In practice, however, the computation becomes very complicated as $n$
grows.
One advantage to consider the algebraic treatment is to give us some
non-trivial implications for the grand partition function as we see in
the next subsection.

\subsection{Implications for the grand partition function}

In the previous subsection, we originally intended to reduce the
dimensions of the phase space integration by commuting the coordinate
$q$ and the momentum $p$.
Though this was not successful because of an anomaly term, the anomaly
term is controlable and gives us an algebraic method to compute 
$\Tr \rho^n$.
In this subsection, we would like to discuss a more important
implication.

Using \eqref{eq:cqcp_comm} and \eqref{eq:cpcq_comm}, it is not difficult
to show
\begin{align}
\Tr(c_qc_p\Pi)^n=i^n\Tr(s_p\Pi^\dagger s_q\Pi)^n
=i^n\Tr(\Pi^\dagger s_q\Pi s_p)^n=i^{2n}\Tr(\Pi^\dagger c_pc_q)^n.
\label{cqcpPi}
\end{align}
Since $\Tr(c_qc_p\Pi)^n$ is real, this should be equal to its
conjugate $\Tr(\Pi^\dagger c_pc_q)^n$.
The observation \eqref{cqcpPi} shows that, if the power $n$ is odd,
we have $\Tr(c_qc_p\Pi)^n=0$ which implies that, by expanding
$\Pi=1-\ket{0_p}\bra{0_q}$, $\Tr\rho^n=\Tr(c_qc_p)^n$ can be expressed
in terms of a certain combination of
$\langle 0_q|(c_qc_p)^m|0_p\rangle$.
The easiest way to incorporate the combinatorics correctly is
as follows.

Let us consider the determinant $\Xi_1(z)=\det (1+z c_p \Pi c_q)$.
This determinant is rewritten as
\begin{align}
\Xi_1(z)
=\det \( 1+z c_p c_q-\frac{z}{4} \ket{0_p}\bra{0_q} \),
\end{align}
where we have used $c_q \ket{0_q}=(1/2) \ket{0_q}$.
This determinant is formally written as
\begin{align}
\Xi_1(z)=\det(1+z c_p c_q)
\det \( 1-\frac{z}{4} \frac{1}{1+z c_p c_q} \ket{0_p}\bra{0_q} \).
\label{eq:Xi1}
\end{align}
The first factor is just the grand partition function $\Xi(z)$, and
(the logarithm of) the second factor is evaluated as
\begin{align}
\log \det\( 1-\frac{z}{4} \frac{1}{1+z c_p c_q} \ket{0_p}\bra{0_q}\)
&=-\Tr \sum_{n=1}^\infty \frac{1}{n}
\( \frac{z}{4} \frac{1}{1+z c_p c_q} \ket{0_p}\bra{0_q} \)^n.
\end{align}
Due to the insertion of the projection operator, the last equation is
drastically simplified
\begin{align}
-\sum_{n=1}^\infty\frac{1}{n} \(\frac{z}{4}\)^n
\( \bra{0_q}\frac{1}{1+z c_p c_q} \ket{0_p} \)^n
=\log \(1-\frac{z}{4}\bra{0_q}\frac{1}{1+z c_p c_q} \ket{0_p} \).
\end{align}
Therefore the relation \eqref{eq:Xi1} becomes
\begin{align}
\Xi_1(z)=\Xi(z)G(-z),
\label{eq:Xi1-Xi}
\end{align}
where we have introduced the new function $G(z)$ by
\begin{align}
G(z)= 1+\frac{z}{4} \bra{0_q}\frac{1}{1-z c_p c_q} \ket{0_p}
=\bra{0_q}\frac{1}{1-z c_q c_p} \ket{0_p}.
\end{align}

Let us show that $\Xi_1(z)$ is invariant under $z \to -z$.
Using the commutation relations \eqref{eq:cqcp_comm} and
\eqref{eq:cpcq_comm}, we have
\begin{align}
\Xi_1(z)=\det (1+z \Pi c_q c_p)
=\det(1+z \Pi i s_p \Pi^\dagger s_q)
=\det (1-z c_p c_q \Pi^\dagger).
\end{align}
Since $\Xi_1(z)$ is a real function for $z \in \mathbb{R}$, this implies
\begin{align}
\Xi_1(z)=\det \bigl(1-z (c_p c_q \Pi^\dagger)^\dagger \bigr)
=\det(1-z \Pi c_q c_p)=\Xi_1(-z).
\label{eq:Xi1-parity}
\end{align}
Plugging \eqref{eq:Xi1-Xi} into \eqref{eq:Xi1-parity}, we find a
relation
\begin{align}
\frac{\Xi(z)}{\Xi(-z)}=\frac{G(z)}{G(-z)}.
\label{eq:XiG}
\end{align}
Now note that the left hand side is nothing but (the square of) the
grand partition function with odd power terms.
More explicitly, let us divide $\Xi(z)$ into two parts
\begin{align}
\Xi(z)=\Xi_{\rm odd}(z) \Xi_{\rm even}(z),
\end{align}
where
\begin{align}
\Xi_{\rm odd}(z)=\exp \left[ - \sum_{n=1}^\infty 
\frac{(-z)^{2n-1}}{2n-1} \Tr \rho^{2n-1} \right],\quad
\Xi_{\rm even}(z)=\exp \left[ - \sum_{n=1}^\infty 
\frac{(-z)^{2n}}{2n} \Tr \rho^{2n} \right].
\label{eq:Xi_odd/even} 
\end{align}
By construction, one easily finds
\begin{align}
\Xi_{\rm odd}(-z)=\frac{1}{\Xi_{\rm odd}(z)},\quad 
\Xi_{\rm even}(-z)=\Xi_{\rm even}(z).
\end{align}
Then, \eqref{eq:XiG} shows
\begin{align}
\Xi_{\rm odd}(z)^2=\frac{\Xi(z)}{\Xi(-z)}=\frac{G(z)}{G(-z)}.
\label{eq:Xi_odd^2}
\end{align}
Using this relation, we can compute $\Tr \rho^{2n+1}$ from $G(z)$.
As we will see in the next section, the function $G(z)$ also has an
important meaning for the even power terms, which connects the two
determinants $\Xi_+(z)$ and $\Xi_-(z)$ defined in the previous
subsection.

After introducing the bra-kets such as $\bra{0_p}$, we can simplify
various definitions and relations appearing previously.
Using the bra-ket notations, the functions $\phi_+^k(q)$ and
$\tilde{\phi}_+^k(q)$ are expressed as
\begin{align}
\phi_+^k(q)=\bra{q} (c_q c_p)^k \ket{0_p},\quad
\tilde{\phi}_+^k(q)=\bra{q} (c_p c_q)^k \ket{0_q}.
\label{eq:phi_+^k-rep}
\end{align}
It is easily shown that $\phi_+^k(q)$ satisfies the recurrence
relation \eqref{eq:phipm_rec}
\begin{align}
\phi_+^k(q)=\int\!\frac{dq'}{2\pi}\bra{q}c_qc_p
\ket{q'}\bra{q'} (c_q c_p)^{k-1} \ket{0_p}
=c_q \int\!\frac{dq'}{2\pi} c_{q-q'}\phi_+^{k-1}(q').
\label{eq:phi_rec}
\end{align}
Similarly, $\tilde{\phi}_+^k(q)$ also satisfies the recurrence relation
\begin{align}
\tilde{\phi}_+^k(q)=\int\! \frac{dq'}{2\pi} c_{q-q'} c_{q'} 
\tilde{\phi}_+^{k-1}(q').
\label{eq:tphi_rec}
\end{align}
Finally let us comment on some useful algebraic relations steming from
\eqref{eq:rho_odd} and \eqref{eq:rho_even} by setting one of the
coordinate to zero.
First, comparing \eqref{eq:phi_+^k-rep} with \eqref{eq:rho^n_ele},
one finds the relation between $\rho^k$ and $\tilde{\phi}_+^k$
\begin{align}
\rho^k(q,0)=\sqrt{2c_q} \tilde{\phi}_+^k(q).
\label{eq:rho-tphi}
\end{align}
Since $\rho_-^k(q,0)=0$ for $k \geq 1$, $\rho^k(q,0)=\rho_+^k(q,0)$
holds.
Secondly, by definition of $G(z)$, the expansion of $G(z)$ is given by
\begin{align}
G(z)=\sum_{n=0}^\infty g_n z^n,\quad g_n=\bra{0_q} (c_q c_p)^n \ket{0_p}. 
\label{eq:G_exp}
\end{align}
From \eqref{eq:phi_+^k-rep} and \eqref{eq:G_exp}, one immediately
finds
\begin{align}
\phi_+^k(0)=g_k \label{eq:phip^k0}.
\end{align}
Thus, with \eqref{eq:rho-tphi} and \eqref{eq:phip^k0},
\eqref{eq:rho_odd} and \eqref{eq:rho_even} imply novel relations
between $\phi_+^k(q)$ and $\tilde{\phi}_+^k(q)$
\begin{align}
\tilde{\phi}_+^{2n+1}(q)&=\frac{1}{4\cosh(\frac{q}{2})} 
\sum_{k=0}^{2n} (-1)^k \phi_+^k(q) g_{2n-k},
\label{eq:tphi_odd}\\
\tilde{\phi}_+^{2n}(q)
&=\frac{\cosh(\frac{q}{2})}{4\sinh^2(\frac{q}{2})} 
\sum_{k=0}^{2n-1} (-1)^k \phi_+^k(q) g_{2n-1-k}.
\label{eq:tphi_even}
\end{align}
Recall that $\phi_+^k(q)$ and $\tilde{\phi}_+^k(q)$ are related
through the Fourier transformation originally.

\section{Exact results}\label{sec:exact}
In this section, we give some exact results for the grand partition
function at $k=1$.
After the study of the algebraic treatment in the previous section, 
our strategy to compute $\Tr \rho^n$ explained in Section
\ref{sec:GPF} is improved.
There are three steps to know $\Tr \rho^n$.
We first fix $\phi_\pm^k(q)$ by using the recurrence relations
\eqref{eq:phipm_rec}.
We then compute $\Tr \rho^n$ separately for odd $n$ and for even $n$.
The computation of $\Tr \rho^{2n+1}$ reduces to the problem to compute
the expansion coefficients $g_n$ of $G(z)$, which are fixed by
$\phi_+^k(q)$.
For the computation of $\Tr \rho^{2n}$, on the other hand, we use the
formula \eqref{eq:rho_even}.
From this prescription, we have determined the exact partition
functions of the ABJM theory at $k=1$ up to $N=9$.

\subsection{Exact partition functions}
Let us start by determining $\phi_\pm^k(q)$.
Since the integral equations \eqref{eq:phipm_rec} have the convolution
integral, the Fourier transformation is useful to solve
\eqref{eq:phipm_rec}.
Let us consider $\phi_+^k(q)$.
The initial condition is $\phi_+^0(q)=1$.
One can trivially find $\phi_+^1(q)=c_q/2$.
Using the Fourier transformation, the convolution between $c_q$ and
$\phi_+^1(q)$ can be evaluated as
\begin{align}
\int\! \frac{dq'}{2\pi}c_{q-q'}\phi_+^1(q')
=\frac{1}{2}\int\! \frac{dp}{2\pi} e^{\frac{i}{2\pi}pq} c_p^2
=\frac{1}{4\pi}qs_q,
\end{align}
where we have used the formula
\begin{align}
c_p=\int\! \frac{dq}{2\pi} e^{-\frac{i}{2\pi}pq}c_q.
\end{align}
Thus we obtain $\phi_+^2(q)=q s_{2q}/(4\pi)$.
Iterating this method, we can fix $\phi_+^k(q)$ recursively.
We list the explicit forms of the first seven functions below
\begin{align}
\phi_+^0(q)&=1,\quad
\phi_+^1(q)=\frac{1}{2}c_q,\quad
\phi_+^2(q)=\frac{1}{4\pi}qs_{2q},\quad
\phi_+^3(q)=\frac{1}{16\pi}c_q(\pi c_q-2qs_{2q}),\nonumber\\
\phi_+^4(q)&=\frac{1}{128\pi^2}
\bigl(4q^2s_{2q}^2+\pi^2(c_q-c_q^2-c_{q/2}^2)\bigr),\nonumber\\
\phi_+^5(q)&=\frac{1}{256\pi^2}c_q
\bigl(4q^2s_{2q}^2+4\pi c_q(qs_{2q}+1)-\pi^2(3c_q^2+c_{q/2}^2)\bigr),\nonumber\\
\phi_+^6(q)&=\frac{1}{384\pi^3}q^3s_q^3c_q^3
+\frac{1}{512\pi}qs_qc_q^3(1+4c_{q/2}^2)
-\frac{1}{1536\pi}(16c_q^3+2c_q^2+c_q-c_{q/2}^2)\nonumber\\
&\quad+\frac{1}{24\pi}(qs_q-1)s_q^2c_q^3
+\frac{1}{1024}c_q^2(1+2c_q).
\label{eq:phi_+}
\end{align}
Similarly $\phi_-^k(q)$ is also fixed.
The explicit forms for $0\leq k \leq 6$ are given by
\begin{align}
\phi_-^0(q)&=1,\quad
\phi_-^1(q)=\frac{1}{2}c_{q/2}^2,\quad
\phi_-^2(q)=\frac{1}{4\pi}c_q(qs_{q}-\pi c_{q/2}^2),\nonumber\\
\phi_-^3(q)&=\frac{1}{16\pi} s_q^2
\bigl(4-2qs_q(1+2c_q)-\pi(1-2c_q)\bigr),\nonumber\\
\phi_-^4(q)&=\frac{1}{32\pi^2}q^2s_{2q}^2
+\frac{1}{32\pi}s_q^2\bigl(4(1-c_q)-2qs_{2q}(1+2c_q)\bigr)
-\frac{1}{128}c_qc_{q/2}^2(6c_q+5),\nonumber\\
\phi_-^5(q)
&=\frac{1}{64\pi^2}
(q^2s_q^2c_q^2c_{q/2}^2+8qs_q^3c_q^2+4qs_qc_q^2c_{q/2}^2-2s_q^2)\nonumber\\
&\quad+\frac{1}{64\pi}
(-4qs_q^3c_q^2-qs_qc_q^2c_{q/2}^2+3s_q^2-4s_q^2c_q)\nonumber\\
&\quad
+\frac{1}{256}s_{q/2}^2
(12c_q^2c_{q/2}^4+8c_qc_{q/2}^4+5c_{q/2}^4-3c_{q/2}^2),\nonumber\\
\phi_-^6(q)
&=\frac{1}{384\pi^3}q^3s_q^3c_q^3
+\frac{1}{128\pi^2}
(-q^2s_q^2c_q^3c_{q/2}^2
+8qs_q^3c_q^3
+4qs_qc_q^3s_{q/2}^2 
+4qs_qc_{q}^3\nonumber\\
&\qquad\qquad\qquad\qquad\qquad
+156c_q^3c_{q/2}^2+26c_q^2c_{q/2}^2-24c_q^3-4c_q^2
+456s_q^2c_{q}^3-120c_q^3s_{q/2}^2)\nonumber\\
&\quad+\frac{1}{1536\pi}
(288s_q^4c_q^2-72s_q^4+52qs_q^3c_q^3-12qs_q^3c_q^2-15qs_q^3c_q
+160s_q^2c_q^2-36s_q^2c_q+20c_q^2)\nonumber\\
&\quad+\frac{1}{1024}
(-32s_q^2c_q^3c_{q/2}^2+224s_q^2c_q^3-54c_q^3s_{q/2}^2
+72c_{q}^3c_{q/2}^2+12c_q^2c_{q/2}^2-6c_q^3-c_q^2).
\label{eq:phi_-}
\end{align}

Now we proceed to computing $\Tr \rho^{2n+1}$.
For this purpose, let us start by computing
$g_n=\bra{0_q}(c_qc_p)^n\ket{0_p}$.
It is easy to see that $g_n$ has the following integral form
\begin{align}
g_n=\int\! \frac{dq}{2\pi} \bra{0_q} (c_q c_p)^m \ket{q}
\bra{q} (c_q c_p)^{n-m} \ket{0_p}
=\int\! \frac{dq}{2\pi} \tilde{\phi}_+^m(q) \phi_+^{n-m}(q).
\end{align}
Note that $g_n$ obviously does not depend on the insertion position
$m$ ($m=0,\cdots,n$), thus we can choose it handily.
If $m=0$ is chosen, we obtain \eqref{eq:phip^k0} because of
$\tilde{\phi}_+^0(q)=2\pi \delta(q)$.
Using the results \eqref{eq:phi_+}, we find the first fourteen values of
$g_n$
\begin{align}
g_0&=1,\quad g_1=\frac{1}{4},\quad g_2=\frac{1}{8\pi},\quad
g_3=\frac{\pi-2}{64\pi},\quad g_4=\frac{1}{2^7 \pi^2},\quad
g_5=\frac{-\pi^2+3\pi+1}{2^9\pi^2},\nonumber\\
g_6&=\frac{9\pi^3-28\pi^2+6}{2^{11}\cdot 3^2 \pi^3},\quad
g_7=\frac{-45\pi^3+92\pi^2+162\pi-12}{2^{14}\cdot 3^2 \pi^3},\nonumber\\
g_8&=\frac{18\pi^3-56\pi^2+3}{2^{15}\cdot 3^2 \pi^4},\quad
g_9=\frac{72\pi^4-219\pi^3-74\pi^2+162\pi+3}
{2^{17}\cdot 3^2\pi^4},\nonumber\\
g_{10}&=\frac{-2475\pi^5+5484\pi^4+7650\pi^3-1400\pi^2+30}
{2^{19}\cdot 3^2 \cdot 5^2 \pi^5}, \nonumber \\
g_{11}&=\frac{4950\pi^5-8684\pi^4-22725\pi^3+1700\pi^2+6075\pi-30}
{2^{22}\cdot 3^2 \cdot 5^2 \pi^5}, \nonumber \\
g_{12}&=\frac{2025\pi^6-57150\pi^4+118312\pi^4+132300\pi^3-8400\pi^2+90}
{2^{23}\cdot 3^4 \cdot 5^2 \pi^6}, \nonumber \\
g_{13}&=\frac{-168075 \pi ^6+493668 \pi ^5+273112 \pi ^4
-522450 \pi ^3-9750 \pi ^2+65610 \pi+90}
{2^{25}\cdot 3^4 \cdot 5^2 \pi^6}.
\end{align}
These lead to the exact value of $\Tr \rho^{2n+1}$ from
\eqref{eq:Xi_odd^2}
\begin{align}
\Tr \rho&=\frac{1}{4},\quad \Tr \rho^3=\frac{\pi-3}{16\pi},\quad
\Tr \rho^5=\frac{10-\pi^2}{256\pi^2},\nonumber\\
\Tr \rho^7&=\frac{-27\pi^3+49\pi^2+126\pi-42}{2^{10}\cdot  3^2 \pi^3},\quad
\Tr \rho^9=\frac{5\pi^4-20\pi^2-96\pi+12}{2^{15}\pi^4},\nonumber\\
\Tr \rho^{11}&=\frac{4725\pi^5-6303\pi^4-25300\pi^3
-12100\pi^2+23100\pi-660}
{2^{17}\cdot 3^2 \cdot 5^2 \pi^5}, \nonumber \\
\Tr \rho^{13}&=\frac{-30375\pi^6+10114\pi^4+655200\pi^3
+978900\pi^2-561600\pi+4680}{2^{21}\cdot 3^4 \cdot 5^2 \pi^6}.
\end{align}

Finally let us see the computation of $\Tr \rho^{2n}$.
We have already determined $\phi_\pm^k(q)$.
Using these results, we can know the matrix elements
$\rho_\pm^k(q_1,q_2)$.
The explicit forms of $\rho_\pm^k(q_1,q_2)$ are given in Appendix
\ref{app:explicit_rhopm}.
These matrix elements enable us to compute $\Tr \rho_\pm^n$.
Here we give the exact values of $\Tr \rho_\pm^{2n}$ ($1 \leq n \leq 4$) 
\begin{align}
\Tr \rho_+^2&=\frac{1}{16\pi},\quad
\Tr \rho_+^4=\frac{\pi^2-8}{512\pi^2},\quad
\Tr \rho_+^6= \frac{9\pi-28}{2^{12}\cdot 3 \pi},\nonumber\\
\Tr \rho_+^8&=\frac{-87\pi^2+192\pi+256}{2^{17}\cdot 3 \pi^2},\label{eq:Trrhop}
\end{align}
and
\begin{align}
\Tr \rho_-^2&=\frac{\pi-3}{16\pi},\quad
\Tr \rho_-^4=\frac{11\pi^2-32\pi-8}{512\pi^2},\quad
\Tr \rho_-^6=\frac{-11\pi^2+4\pi+96}{2^{12} \pi^2},\nonumber\\
\Tr \rho_-^8&=\frac{-1467\pi^3+3520\pi^2+5376\pi-6144}
{2^{17}\cdot 3^2 \pi^3}.\label{eq:Trrhom}
\end{align}
Note that $\Tr \rho^n$ is computed by using \eqref{eq:Trrho^n_split}.

Collecting the above results, the grand partition function is expanded
as
\begin{align}
\Xi (z)&=
1+\frac{z}{2^2}+\frac{z^2}{2^4\pi}
+\frac{\pi-3}{2^6\pi}z^3+\frac{10-\pi^2}{2^{10}\pi^2}z^4
+\frac{-9\pi^2+20\pi+26}{2^{12}\pi^2}z^5
+\frac{36\pi^3-121\pi^2+78}{2^{14} \cdot 3^2\pi^3}z^6\notag\\
&\quad
+\frac{-75\pi^3+193\pi^2+174\pi-126}{2^{16}\cdot 3 \pi^3}z^7
+\frac{1053\pi^4-2016\pi^3-4148\pi^2+876}{2^{21}\cdot 3^2 \pi^4}z^8\nonumber\\
&\quad
+\frac{5517\pi^4-13480\pi^3-15348\pi^2+8880\pi+4140}{2^{23}\cdot 3^2 \pi^4}z^9
+{\cal O}(z^{10}).
\end{align}
Let us recall that the coefficient of $z^N$ in $\Xi(z)$ is just the
partition function $Z(N)$, thus we obtain the exact ABJM partition
function $Z(N)$ at $k=1$ up to $N=9$.
The partition functions for higher $N$ can also be computed in this
way.

\subsection{Novel relation between $G(z)$ and $\Xi_\pm (z)$}
From the results in the previous subsection, we find a novel relation
\begin{align}
\log G(z)=\sum_{n=1}^\infty \frac{z^{2n-1}}{2n-1}\Tr \rho^{2n-1}
+\sum_{n=1}^\infty \frac{z^{2n}}{2n}(\Tr\rho_+^{2n}-\Tr\rho_-^{2n})
+{\cal O}(z^{10}),
\end{align}
in the expansion around $z=0$.
As we will see below, it is shown that the above relation holds
exactly for all $z$, that is,
\begin{align}
G(z)=\frac{\det(1+z\rho_-)}{\det(1-z\rho_+)}
=\frac{\Xi_-(z)}{\Xi_+(-z)}.
\label{eq:G-Xipm}
\end{align}

Before proceeding to the proof, let us comment on a consequence led by
the above equality.
Using \eqref{eq:G-Xipm}, the grand partition function is written as
\begin{align}
\Xi(z)=\det (1-z^2 \rho_+^2)G(z).
\end{align}
Since the expansion coefficients of $G(z)$ are just given by
$\phi_+^k(0)$ (see \eqref{eq:phip^k0}), the function $G(z)$ is fixed
by $\phi_+^k(0)$.
On the other hand, the matrix elements $\rho_+^{2n}(q_1,q_2)$ are
fixed by $\phi_+^k(q)$ through \eqref{eq:rho_even}.
Therefore $\det (1-z^2 \rho_+^2)$ is also determined by
$\phi_+^k(q)$.
We conclude that the total grand partition function is completely
fixed by $\phi_+^k(q)$ (or $\rho_+$) only!
We do not need any information about $\phi_-^k(q)$ (or $\rho_-$) to
know $\Xi(z)$.
The novel relation we have found in this section suggests that we have
an alternative strategy for computing the grand partition function.
Previously we need to compute $\Tr \rho^{n}$. 
In particular, to compute $\Tr \rho^{2n}$, we need both of $\Tr \rho_\pm^{2n}$ and sum up
them: $\Tr \rho^{2n}=\Tr \rho_+^{2n}+\Tr \rho_-^{2n}$.
After knowing the relation, the ingredients we need are  
the coefficients $g_n$ of $G(z)$ and $\Tr \rho_+^{2n}$.
$\Tr \rho_-^{2n}$ is automatically fixed by the above relation.

The rest of this subsection is devoted to the proof of
\eqref{eq:G-Xipm} by using the properties of the parity-preserving
Hankel matrix $M$.
Using the relation \eqref{eq:Xi_odd^2}, one can check that the odd
power part of \eqref{eq:G-Xipm} holds for all-order in $z$.
Thus we concentrate ourselves to the even power part.
The statement to be shown is now rephrased as
\begin{align}
G(z)G(-z)=\frac{\det(1-z^2\rho_-^2)}{\det(1-z^2\rho_+^2)}.
\label{eq:G-Xipm-2}
\end{align}
Let us recall that $M$ decomposes into two Hankel matrices $M_\pm$.
From \eqref{eq:Mm-Mp}, we find
\begin{align}
M_-^2=M_+T_-T_+M_+=M_+P M_+,
\end{align}
where $P =1- \ket{e_0}\bra{e_0}$ and $\ket{e_0}\bra{e_0}$ is the
projector to the zero-th component.
This relation implies
\begin{align}
\Tr M_-^{2n}=\Tr(M_+^2P)^n
\end{align}
The right hand side looks similar to \eqref{cqcpPi} and therefore
suggests that a similar manipulation is possible.
All we have to do is to rewrite the right hand side by expanding
$P=1-\ket{e_0}\bra{e_0}$.

Again, this is most easily done by studying the determinants.
Let us rewrite the right hand side of \eqref{eq:G-Xipm-2}.
We use the fact $\Xi_\pm(z)=\det(1+z M_\pm)$, then
\begin{align}
\frac{\det(1-z^2M_-^2)}{\det(1-z^2M_+^2)}
=\frac{\det(1-z^2M_+^2 P)}{\det(1-z^2 M_+^2)}
=\det\(1+\frac{z^2 M_+^2}{1-z^2 M_+^2}\ket{e_0}\bra{e_0}\).
\end{align}
From the similar argument to derive \eqref{eq:Xi1-Xi}, this
determinant becomes simple
\begin{align}
\det\(1+\frac{z^2 M_+^2}{1-z^2 M_+^2}\ket{e_0}\bra{e_0}\)
=1+\bra{e_0}\frac{z^2 M_+^2}{1-z^2 M_+^2}\ket{e_0}
=\bra{e_0}\frac{1}{1-z^2 M_+^2}\ket{e_0}.
\end{align}
Expanding this in $z$, we obtain
\begin{align}
\frac{\det(1-z^2M_-^2)}{\det(1-z^2M_+^2)}
=\sum_{n=0}^\infty z^{2n} (M_+^{2n})_{00}.
\label{eq:ratio}
\end{align}
Now we want to write the right hand side of \eqref{eq:ratio} in terms
of the coefficients $g_n$ of $G(z)$.
To do this, we use \eqref{eq:phi-Hankel} and \eqref{eq:tphi_odd}. 
Setting $q=0$ in the second equation of \eqref{eq:phi-Hankel}, we
obtain
\begin{align}
4\tilde{\phi}_+^k(0)=(M_+^{k-1})_{00}.
\end{align}
On the other hand, if setting $q=0$ in \eqref{eq:tphi_odd}, we also
obtain
\begin{align}
4\tilde{\phi}_+^{2n+1}(0)=\sum_{m=0}^{2n} (-1)^m g_m g_{2n-m}.
\end{align}
where we have used \eqref{eq:phip^k0}.
Comparing these relations, we find
\begin{align}
(M_+^{2n})_{00}=\sum_{m=0}^{2n} (-1)^m g_m g_{2n-m}.
\label{eq:M+^2n_00}
\end{align}
Substituting \eqref{eq:M+^2n_00} into \eqref{eq:ratio}, one can
confirm that \eqref{eq:G-Xipm-2} really holds.

\section{Numerical results}
In the previous section, we compute the (grand) partition function
analytically.
Here we discuss another approach to analyze the partition function of
the ABJM theory.
The key idea is to use a link between the Fredholm theory for the
kernel like \eqref{eq:rho} and TBA system.
This novel link was first considered by Zamolodchikov to study
two-dimensional polymers in \cite{Z}.
The kernel appearing there is similar to \eqref{eq:rho}.
Thus we can use some of the results in \cite{Z}.
Particularly, we can write down the TBA-type integral equations that
give $\Tr \rho^n$.
Though the TBA equations can not be solved analytically in general, it
is easy to solve them at least numerically.
In this section, we numerically solve the TBA-like equations and
compute $\Tr \rho^n$.
We compare them with the exact results in the previous section.
Our numerical results show the very good agreement with the exact
ones, and further give a prediction of $\Tr \rho^n$ (and also the
partition function), of which we have not found the exact values, with
a very high degree of accuracy.
We also study numerically the eigenvalue spectrum of the density
matrix.
 
\subsection{Partition function from TBA-like equations}
In order to see the relation to the TBA-like equations, it is
convenient to rewrite the kernel \eqref{eq:rho} as the form
\begin{align}
\rho (q_1, q_2)
=\frac{e^{-\frac{1}{2}U(q_1)-\frac{1}{2}U(q_2)}}{2 \cosh(\frac{q_1-q_2}{2k})},
\label{eq:rho2}
\end{align}
where
\begin{align}
U(q)=\log \left[ 2\cosh \frac{q}{2} \right].
\label{eq:U}
\end{align}
In \cite{Z}, Zamolodchikov considered the following integral
\begin{align}
R_n(x)=e^{-2u(x)}\mint dx_1 \cdots dx_n
\frac{e^{-2u(x_1)-\cdots-2u(x_n)}}{\cosh(\frac{x-x_1}{2})\cosh(\frac{x_1-x_2}{2})
\cdots \cosh(\frac{x_n-x}{2})}.
\end{align}
One can easily check that if $u(x)=U(kx)/2$ is chosen, the integral of
$R_n(x)$ is related to $\Tr \rho^{n+1}$ thorough
\begin{align}
\int_{-\infty}^\infty dx\, R_n(x)=(4\pi)^{n+1}\Tr \rho^{n+1}.
\label{eq:R-Z}
\end{align}
Let us introduce the generating functionals of $R_n(x)$,
\begin{align}
R_e(x|z)&=\sum_{m=0}^\infty \(-\frac{z}{4\pi}\)^{2m} R_{2m}(x),\quad
R_o(x|z)=\sum_{m=0}^\infty\(-\frac{z}{4\pi}\)^{2m+1}R_{2m+1}(x),
\label{eq:exp_R_eo}\\
R(x|z)&=R_e(x|z)+R_o(x|z).
\end{align}
From \eqref{eq:GPF} and \eqref{eq:R-Z}, we obtain ($z=e^\mu$)
\begin{align}
\frac{1}{4\pi}\int_{-\infty}^\infty dx\, R(x|z)
=\frac{d}{dz} \log \Xi(z)
=e^{-\mu}J'(\mu),
\label{eq:int_R}
\end{align}
where $J(\mu)=\log \Xi(z)$ is the grand potential.
Thus if we know the function $R(x|z)$, we can obtain the grand
partition function.

One of the important results in \cite{Z} is that the function
$R_e(x|z)$ is conjectured to be the solution of the following TBA-type
integral equations
\begin{align}
\log R_e(x|z)&=\log s(kx)+\int_{-\infty}^\infty dx'\,
K_1(x-x')\log(1+\eta^2(x'|z)), \label{eq:int_eq1}\\
\eta (x|z)&=-z \int_{-\infty}^\infty dx'\,
K_1(x-x') R_e(x'|z), \label{eq:int_eq2}
\end{align}
where $\eta(x|z)$ is an auxiliary function, and we have introduced the
following functions
\begin{align}
s(x)=\frac{1}{2\cosh \frac{x}{2}},\quad
K_1(x)=\frac{1}{2\pi \cosh x}.
\label{eq:kernel0}
\end{align}
These two equations determine the functions $R_e(x|z)$ and $\eta(x|z)$
uniquely for given $z$.
Once $R_e(x|z)$ and $\eta(x|z)$ are determined, the function
$R_o(x|z)$ is given by
\begin{align}
R_o(x|z)=R_e(x|z)\int_{-\infty}^\infty dx'\, K_2(x-x') \arctan \eta(x'|z),
\label{eq:int_eq3}
\end{align}
where
\begin{align}
K_2(x)=\frac{1}{\pi \cosh^2 x}.
\end{align}
Therefore we can obtain the whole function $R(x|z)$ from these
integral equations.
In our case, the driving term in \eqref{eq:int_eq1} takes the form
$-\log (2\cosh (kx/2))$, which is not a standard form $-m\cosh x$ in
ordinary TBA systems for 2d relativistic quantum field theories.
Note that this conjecture has been proved in \cite{TW}.

\subsection{Numerical results from TBA}
Now we want to solve the TBA-like equations \eqref{eq:int_eq1},
\eqref{eq:int_eq2} and \eqref{eq:int_eq3} and to compute
$\Tr \rho^n$.
For this purpose, we expand $\eta(x|z)$ as
\begin{align}
\eta(x|z)=\sum_{m=0}^\infty \( -\frac{z}{4\pi}\)^{2m+1} \eta_{2m+1}(x).
\label{eq:exp_eta}
\end{align}
Substituting \eqref{eq:exp_R_eo} and \eqref{eq:exp_eta} into
\eqref{eq:int_eq1}, we obtain the following first few equations,
\begin{align}
&\log R_0(x)=\log s(kx), \quad 
\frac{R_2(x)}{R_0(x)}=K_1* \eta_1^2, \nonumber\\
&\frac{R_4(x)}{R_0(x)}-\frac{1}{2}\( \frac{R_2(x)}{R_0(x)} \)^2
=K_1*\( 2\eta_1\eta_3-\frac{\eta_1^4}{2}\),
\label{eq:Reven}
\end{align}
where the convolution is defined by
\begin{align}
f*g=\int_{-\infty}^\infty dx' f(x-x')g(x').
\end{align}
Similarly, from \eqref{eq:int_eq2}, we obtain
\begin{align}
\eta_{2m+1}(x)=4\pi K_1* R_{2m},
\label{eq:eta_odd}
\end{align}
for $m=0,1,\cdots$.
Starting with $R_0(x)=s(kx)$, one can determine $\eta_1(x)$ from
\eqref{eq:eta_odd}, then $R_2(x)$ is determined by the second equation
in \eqref{eq:Reven}.
The flow of the determined functions is as follows,
\begin{align}
R_0(x) \to \eta_1(x) \to R_2(x) \to \eta_3(x) \to R_4(x) \to \cdots.
\end{align}
In the similar way, $R_n(x)$ for odd $n$ is fixed from
\eqref{eq:int_eq3} such as
\begin{align}
R_1(x)=R_0(x)K_2*\eta_1,\quad
R_3(x)=R_0(x) K_2*\( \eta_3-\frac{\eta_1^3}{3}\)+R_2(x) K_2*\eta_1.
\label{eq:R1&R3}
\end{align}

The integral equations \eqref{eq:Reven}, \eqref{eq:eta_odd},
\eqref{eq:R1&R3} etc. are solved recursively.
In principle, we can solve them analytically by evaluating the
convolutions of the known functions at each step.
Technically, however, it is very messy to do that.
Alternatively, here we solve these sequential integral equations
numerically at $k=1$.\footnote{%
We have also analytically checked that the TBA-like equations indeed
reproduce the correct $\Tr \rho^n$ up to $n=5$.}
If $R_n(x)$ is determined, we can compute $\Tr \rho^{n+1}$ from
\eqref{eq:R-Z}.

We have indeed computed the numerical values of $\Tr \rho^{n}$ from
such integral equations.
In Table \ref{tab:Zn}, we summarize the numerical values for
$2\leq n \leq 16$.
The numerical results show the very good agreement with the exact
values obtained in the previous section.
We would like to emphasize that the TBA-like equations are valid for general $k$
(even for non-integer values).
The only difference is the initial condition $R_0(x)=s(kx)$.
We can predict the high-accurate values of the partition functions for
various $N$ and $k$ in the same manner.

In this subsection, we expand all the unknown functions $R_e(x|z)$, $R_o(x|z)$ and $\eta(x|z)$ 
in $z$ to compute $\Tr \rho^n$.
We note that the TBA-like equations  \eqref{eq:int_eq1}, \eqref{eq:int_eq2} and \eqref{eq:int_eq3} 
can be also solved for given $z$ by iteration as we will see in the next section.

\begin{table}[tb]
 \caption{Numerical values of $(2\pi)^n \Tr \rho^n$
 at $k=1$.
 These numerical values are computed by using the TBA-like equations
 \eqref{eq:int_eq1}, \eqref{eq:int_eq2} and \eqref{eq:int_eq3}. Our numerical results nicely agree
 with the exact values in the previous section.}
 \label{tab:Zn}
 \begin{center}
  \begin{tabular}{ccc}
    \hline
     $n$  &  $(2\pi)^n \Tr \rho^n$  
 & Errors \\
    \hline
      2 &  0.896604773477443035477301058329286341729730
 & $1.1 \times 10^{-40}$ \\
      3 &  0.698731738515872159486421033736470898141922
 & $1.7 \times 10^{-40}$ \\
      4 &  0.587330256906414478771564191224945948765499
 & $2.1 \times 10^{-40}$ \\
      5 &  0.505385252214593561502730078447293192743804
 & $2.2 \times 10^{-40}$ \\
      6 &  0.438346955317146787321292739552212309497233
 & $2.3 \times 10^{-40}$ \\
      7 &  0.381274926859615816279007202371073982670788
 & $2.2 \times 10^{-40}$ \\
      8 &  0.331972050713896807160802466534622236608778
 & $2.2 \times 10^{-40}$ \\
      9 &  0.289152279521382289292363666349330742224258
 & $2.0 \times 10^{-40}$ \\
      10 & 0.251890109244736461833445267225791497711006
 & \\
      11 & 0.219440841867035365985490063842951367327288
 & $1.7 \times 10^{-40}$ \\
      12 & 0.191175332952772245351675601074802777111664
 & \\
      13 & 0.166551755978085254984697038651649345830601
 & $1.4 \times 10^{-40}$ \\
      14 & 0.145100086034249146678243847815416397520091
 & \\
      15 & 0.126411483244355445082928632866179620259652
 & \\
      16 & 0.110129973053738524406779285719767961758381
 & \\
    \hline
  \end{tabular}
 \end{center}
\end{table}

\subsection{Eigenvalue distribution}\label{sec:num_eigen}
Here we study the eigenvalue distribution of the density matrix
numerically.
For this purpose, we use the Hankel matrix representation discussed in
Section \ref{sec:eigen}.
Since this matrix is discrete, we can evaluate its eigenvalues by the
level truncation.
Furthermore the Hankel matrix elements $M_{ij}$ only depend on $i+j$.
This means that we can reduce the ${\cal O}(L^2)$-computation into
${\cal O}(L)$-computation where $L$ is the size of the matrix. 
This is the reason why we use the Hankel matrix $M$ rather than the
Chebyshev representation $M^\rho$ of the density matrix.
Physically, we are interested in the energy spectrum of the Fermi gas
system.
Since the Hamiltonian is related to the density matrix as
$\rho=e^{-H}$, the $n$-th energy eigenvalue is given by
\begin{align}
E_n=-\log \lambda_n,
\end{align}
where $\lambda_n$ is the eigenvalue of the density matrix.

\begin{table}[tbp]
 \caption{Numerical energy eigenvalues of the Fermi gas system at $k=1$. 
 The values at $\infty$ are obtained by extrapolating the data with
 the functions $e^{(0)}+e^{(1)}/L+e^{(2)}/L^2+e^{(3)}/L^3$ where $L$
 is the size of the matrix.
 $E_{\pm,n}$ mean the $n$-th energy eigenvalues of the systems
 described by the density matrices $\rho_\pm$.}
 \label{tab:eigen}
 \begin{center}
  \begin{tabular}{cccccccc}
  \hline
     Size & $E_0$ ($E_{+,0}$)   &  $E_1$ ($E_{-,0}$)  & $E_2$ ($E_{+,1}$)   &  
     $E_3$ ($E_{-,1}$)  &  $E_4$ ($E_{+,2}$)  &  $E_5$ ($E_{-,2}$) & $E_6$ ($E_{+,3}$) \\
  \hline
     $100$  &  $1.975939$  &  $2.977194$  &  $3.739240$
  &  $4.432293$  &  $5.142744$  &  $5.895999$  & $6.709097$ \\
     $200$  &  $1.975808$  &  $2.974644$  &  $3.721872$
  &  $4.372444$  &  $5.003785$  &  $5.656839$  & $6.350546$ \\
     $400$  &  $1.975772$  &  $2.973812$  &  $3.715146$
  &  $4.344395$  &  $4.928247$  &  $5.510457$  & $6.116287$ \\
     $1000$  & $1.975760$  &  $2.973523$  &  $3.712410$
  &  $4.330522$  &  $4.883362$  &  $5.408403$  & $5.933915$ \\
     $2000$  & $1.975759$  &  $2.973474$  &  $3.711873$
  &  $4.327247$  &  $4.870444$  &  $5.372886$  & $5.860046$ \\
     $4000$  & $1.975758$  &  $2.973460$  &  $3.711711$
  &  $4.326134$  &  $4.865368$  &  $5.356542$  & $5.820704$ \\
     $10000$ & $1.975758$  &  $2.973456$  &  $3.711658$
  &  $4.325733$  &  $4.863290$  &  $5.348715$  & $5.798487$ \\
     $\infty$  &  $1.975758$  &  $2.973449$  & $3.711541$
  &  $4.324808$  &  $4.859445$  &  $5.339096$  & $5.782347$ \\
  \hline
  \end{tabular}
 \end{center}
\end{table}

\begin{figure}[tbp]
  \begin{center}
    \includegraphics[keepaspectratio=true,height=70mm]{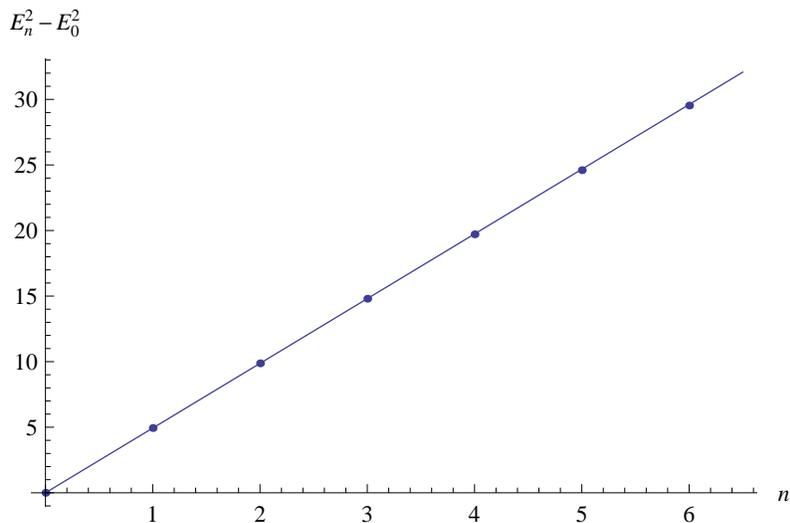}
  \end{center}
  \caption{The (square of) energy spectrum of the Fermi gas system
  against the excitation level $n$.
  The dots correspond to the extrapolated numerical eigenvalues in
  Table \ref{tab:eigen} while the solid line is the plot of
  $E^2-E_0^2=\pi^2 n/2$.}
  \label{fig:Eigen_dis}
\end{figure}

In Table \ref{tab:eigen}, we summarize the numerical result of the
energy eigenvalues of this quantum mechanical system, computed from
the Hankel matrix $M$ using the level truncation.
In Figure \ref{fig:Eigen_dis}, we show the distribution of these
energy eigenvalues.
We have used the extrapolated values as in Table \ref{tab:eigen} to
plot this graph.
We observe that the square of the energy eigenvalue $E_n$ shows the
good linear behavior on the excitation level $n$ for small $n$:
\begin{align}
E_n^2-E_0^2 \approx \frac{\pi^2}{2}n.
\end{align}
Surprisingly, the coefficient of $n$ in this equation agrees with the
one appearing in the relation between the number of states and the
energy in the thermodynamic limit \cite{MPfree}: $n(E)=\frac{2}{\pi^2 k}E^2+n_0$.

Let us compare these results with the estimations from the exact
results in the previous section.
We can estimate the ground state and the first excited state energies
from $\Tr \rho_{\pm}^n$ because 
\begin{align}
\Tr \rho_{\pm}^n =\sum_{\ell=0}^\infty \lambda_{\pm,\ell}^n \approx \lambda_{\pm,0}^n,
\end{align}
for large $n$.
Using \eqref{eq:Trrhop} and \eqref{eq:Trrhop_odd}, we find
\begin{align}
E_0=E_{+,0} \approx -\log\(\frac{\Tr \rho_+^8}{\Tr \rho_+^7}\)
=1.9757623\cdots,
\end{align}
which matches the estimation from the Hankel matrix in Table
\ref{tab:eigen}.
The order of the subleading corrections is
${\cal O}(\lambda_2^7/\lambda_0^7) \sim {\cal O}(10^{-6})$.
Similarly,
\begin{align}
E_1=E_{-,0}\approx -\log \(\frac{\Tr \rho_-^8}{\Tr \rho_-^7}\)
=2.9735127\cdots,
\end{align}
where we have used the exact results \eqref{eq:Trrhom} and
\eqref{eq:Trrhom_odd}.
The result is again consistent with the value in Table
\ref{tab:eigen}.
The subleading corrections come from
${\cal O}(\lambda_3^7/\lambda_1^7) \sim {\cal O}(10^{-5})$.

\section{Comparison with perturbative results}
In this section, we compare our results with the perturbative ones in
\cite{MPfree}.

We first consider the free energy of the ABJM theory.
In \cite{FHM,MPfree}, it was shown that the ABJM partition function is
written in terms of the Airy function if one neglects the instanton
corrections.
Since we have obtained the exact partition function $Z(N)$ up to $N=9$
in Section \ref{sec:exact}, we can compare them with the perturbative
results.
This comparison allows us to know the behavior of the non-perturbative
effects.
Our results suggest that the leading non-perturbative correction scales as $e^{-2\pi\sqrt{2N}}$.

We next consider the expectation value of $N$ (or the derivative of
the grand potential).
Using the TBA-like equations in the previous section, we can
numerically compute $\ev{N}$ as a function of the chemical potential
$\mu$.
Our numerical results show the good agreement with the perturbative
solution in \cite{MPfree} in the large $\mu$ limit.

\subsection{Free energy}
\begin{figure}[tb]
\begin{center}
\begin{tabular}{cc}
\hspace{-3mm}
\resizebox{80mm}{!}{\includegraphics{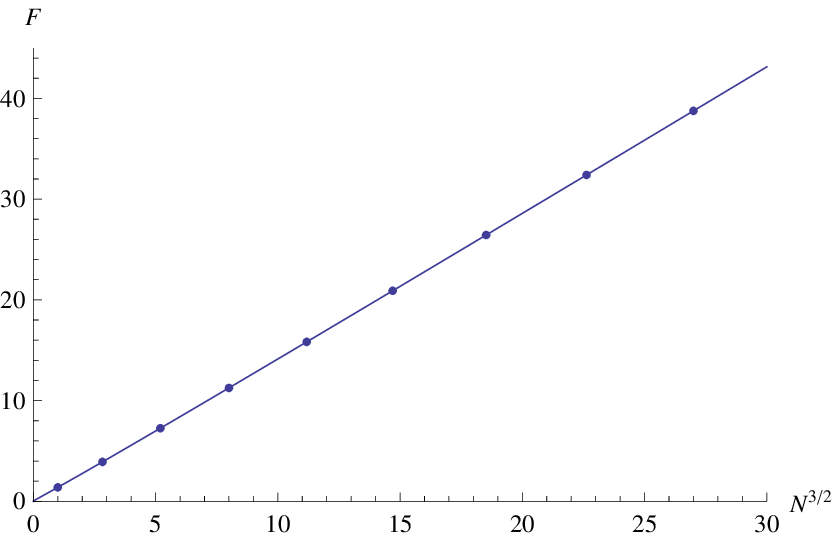}}
\hspace{-4mm}
&
\resizebox{80mm}{!}{\includegraphics{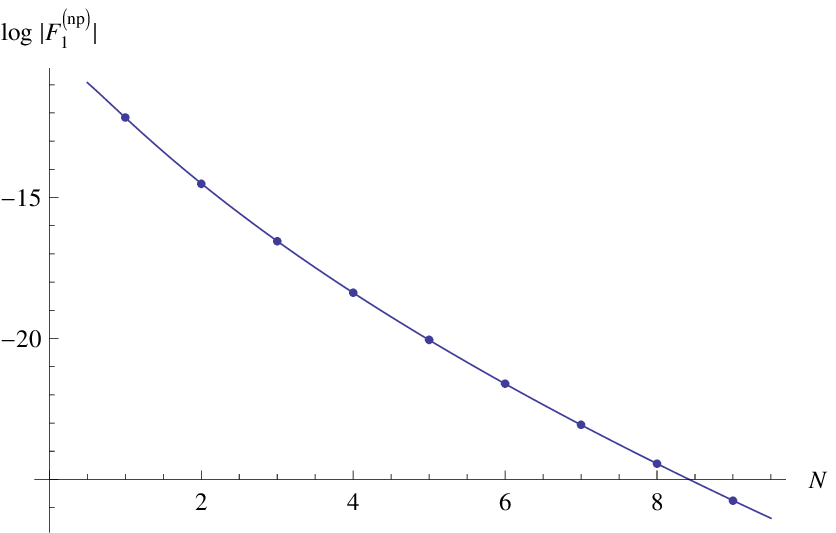}}
\hspace{-5mm}
\\ (a) & (b) 
\vspace{-5mm}
\end{tabular}
\end{center}
  \caption{The behavior of the free energy at $k=1$.
  (a) The dots show our exact values while the solid line shows the
  free energy computed from \eqref{eq:ZMP2}.
  (b) The (log of) non-perturbative correction $F_1^{\rm (np)}(N)$. The dots are the exact data, and
  the solid line is the fitted function \eqref{eq:fitting} with \eqref{eq:fitting_value}.}
  \label{fig:Free_energy}
\end{figure}

Let us consider the free energy of the ABJM theory.
As was shown in \cite{FHM,MPfree}, the partition function is written
in terms of the Airy function if one neglects the instanton
corrections,
\begin{align}
Z_k(N)&=Z_k^{(\rm Airy)}(N)+Z_k^{\rm (np)}(N),\label{eq:ZMP1}\\
Z_k^{(\rm Airy)}(N)&=\(\frac{2}{\pi^2 k}\)^{-1/3}e^{A(k)} 
\Ai \left[ \(\frac{2}{\pi^2 k}\)^{-1/3}
\(N-\frac{1}{3k}-\frac{k}{24} \) \right],
\label{eq:ZMP2}
\end{align}
where $Z_k^{\rm (np)}(N)$ is the non-perturbative contribution.
The function $A(k)$ has the following small $k$ expansion,
\begin{align}
A(k)=\frac{2\zeta(3)}{\pi^2 k}
-\frac{k}{12}-\frac{\pi^2 k^3}{4320}+{\cal O}(k^5).
\end{align}
In \cite{HHHNSY}, it was conjectured that this function has the
following integral expression, which should be valid for arbitrary
$k$,
\begin{align}
A(k)&=\frac{1}{2}\log 2-\frac{\zeta(3)}{8\pi^2}k^2
-\frac{1}{6}\log k +\frac{1}{6}\log \( \frac{\pi}{2}\)+2\zeta'(-1)
\nonumber\\
&\quad
-\frac{1}{3}\int_0^\infty dx \, 
\frac{1}{e^{kx}-1}\(\frac{3}{x^3}-\frac{1}{x}-\frac{3}{x \sinh^2x} \).
\end{align}
Figure \ref{fig:Free_energy} (a) shows the behavior of the free energy
$F(N)=-\log Z(N)$ at $k=1$.
The result \eqref{eq:ZMP2} shows the very close behavior of the exact
results even though the non-perturbative corrections are dropped.
It is surprising that these corrections at $k=1$ is very small even
for small $N$.

Let us consider the non-perturbative corrections
to the free energy
\begin{align}
F_1^{\rm (np)}(N) \equiv F_1^{(\rm exact)}(N)-F_1^{({\rm Airy})}(N),
\end{align}
where $F_k^{({\rm Airy})}(N)=-\log Z_k^{({\rm Airy})}(N)$.%
\footnote{In the definition here,
$F_k^{\rm (np)}(N) \ne -\log Z_k^{\rm (np)}(N)$ where
$Z_k^{\rm (np)}(N)=Z_k^{\rm (exact)}(N)-Z_k^{\rm (Airy)}(N)$.}
We find $F_1^{\rm (np)}(N)$ is negative for all $N\leq 9$. 
In \cite{DMP2,MPfree} it was noticed that there are two kinds of
non-perturbative instanton effects contributing to the free energy.
One of them is the worldsheet instanton $\exp(-2\pi\sqrt{2N/k})$, while
the other is the membrane instanton $\exp(-\pi\sqrt{2kN})$ which comes
from D2-branes wrapping on ${\mathbb RP}^3$ in ${\mathbb CP}^3$.
Since in the case $k=1$ there is no ${\mathbb RP}^3$ to wrap, we
expect that the membrane instanton effect vanishes and the non-perturbative
corrections start from $\exp(-2\pi\sqrt{2N})$.

In Figure \ref{fig:Free_energy} (b), we tried to fit our exact data of $\log|F^{\rm (np)}_1(N)|$ by the function
\begin{align}
\log|F^{\rm (np)}_1(N)|\sim a+b\log N-c\sqrt{N},
\label{eq:fitting}
\end{align}
and found
the best values of the fitting
\begin{align}
a=-3.17,\quad b=2.01,\quad c=9.00~.
\label{eq:fitting_value}
\end{align}
Note especially that the value of $c$ is very close to $2\pi\sqrt{2}=8.88577$. This result
strongly suggests
that the membrane instanton $e^{-\pi\sqrt{2kN}}$
is absent
when $k=1$, and 
the leading non-perturbative correction behaves as
$F^{\rm np}_1(N)\sim e^{-2\pi\sqrt{2N}}$ at large $N$.

Finally, let us briefly comment on the pre-factor of the instanton correction 
$e^{-2\pi\sqrt{2N}}$.
From the general argument in \cite{MPfree}, one expects that
$F_1^{\rm (np)}(N)$ scales as
\begin{equation}
F_1^{\rm (np)}(N)\sim (AN+B \sqrt{N}+C)e^{-2\pi\sqrt{2N}}
\label{eq:Fnp-expected}
\end{equation} 
with some coefficients $A,B$ and $C$. 
Our analysis \eqref{eq:fitting_value}
seems to indicate that the leading term of $F_1^{\rm (np)}(N)$
behaves as $N^2e^{-2\pi\sqrt{2N}}$ rather than $Ne^{-2\pi\sqrt{2N}}$.
However, our result up to $N=9$ is insufficient to tell.
One possibility is that $F_1^{\rm (np)}(N)$ may actually scale as \eqref{eq:Fnp-expected}
for much larger $N$ than $N=9$.
It would be interesting to study the behavior of the non-perturbative correction $F_1^{\rm (np)}(N)$ 
in more detail.

\subsection{Expectation value of $N$}
In the previous section, we solved the TBA-like equations by expanding
all the unknown functions in $z$ in order to compute $\Tr \rho^n$.
It is possible to solve them for given $z$ by iteration.
Let us compare the results from such integral equations with the
perturbative ones as functions of $z=e^\mu$.
For this purpose, it is convenient to focus on the derivative of the
grand potential $J(\mu)$ because this quantity naturally appears in
the TBA approach as in \eqref{eq:int_R}.
From the standard argument of statistical mechanics, the derivative
$J'(\mu)$ corresponds to the expectation value of $N$ in the grand
canonical ensemble
\begin{align}
\ev{N}=J'(\mu)=\frac{e^{\mu}}{4\pi}\int_{-\infty}^\infty dx R(x|e^\mu),
\label{eq:<N>}
\end{align}
where \eqref{eq:int_R} has been used in the last equality.
Using \eqref{eq:rho}, the grand potential is also expressed in terms
of $\Tr \rho^n$.
The expectation value $\ev{N}$ is thus written as the expansion around
$\mu=-\infty$
\begin{align}
\ev{N}=-\sum_{n=1}^\infty (-e^\mu)^n \Tr \rho^n.
\label{eq:<N>_exp}
\end{align}
Let us consider the perturbative contribution of $J(\mu)$.
As shown in \cite{MPfree}, the grand potential of the ABJM theory is
given by
\begin{align}
J^{({\rm pert})}(\mu)=\frac{2\mu^3}{3k\pi^2}+\mu\(\frac{1}{3k}+\frac{k}{24}\)+A(k),
\end{align}
if non-perturbative corrections are neglected.
Thus the derivative of $J(\mu)$ is given by
\begin{align}
\ev{N}^{({\rm pert})}=\frac{2\mu^2}{k\pi^2}+\frac{1}{3k}+\frac{k}{24}.
\label{eq:<N>_pert}
\end{align}
Note that, in \eqref{eq:<N>_pert}, the contribution from $A(k)$ is
dropped.

\begin{figure}[tb]
  \begin{center}
    \includegraphics[keepaspectratio=true,height=80mm]{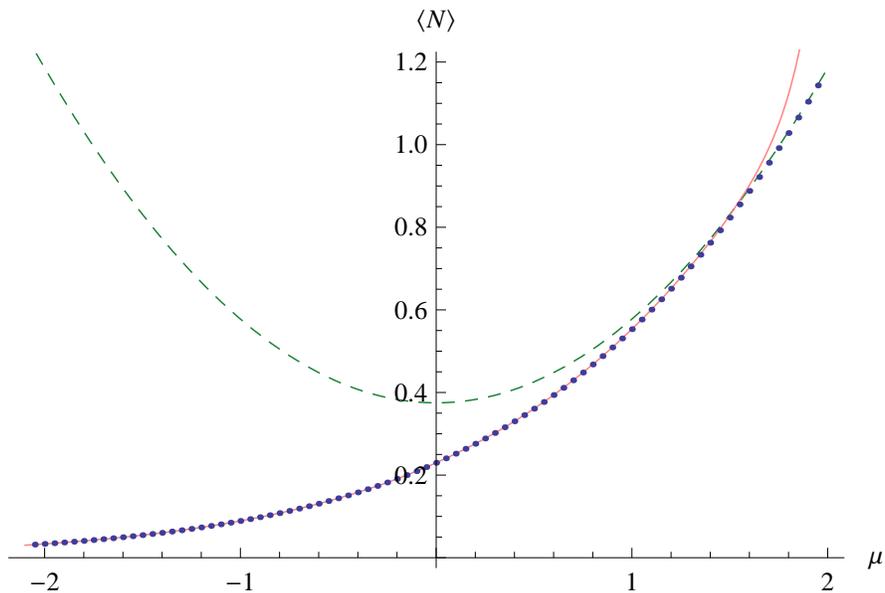}
  \end{center}
  \caption{The behavior of the expectation value of $N$ as a function
  of $\mu$.
  The dots show the results obtained from TBA-like equations.
  The red solid line represents the finite-truncated expansion
  \eqref{eq:<N>_exp} up to $n=9$ with the exact $\Tr \rho^n$ while the
  green dashed line represents the perturbative result
  \eqref{eq:<N>_pert}.}
  \label{fig:mu-N}
\end{figure}

We numerically solve the TBA-like equations \eqref{eq:int_eq1},
\eqref{eq:int_eq2} and \eqref{eq:int_eq3} at $k=1$ for various values
of $z=e^\mu$, and compute the expectation value of $N$ by using
\eqref{eq:<N>} as a function of $\mu$.
The results are shown in Figure \ref{fig:mu-N}.
In the figure, the dots represent our numerical results from the
TBA-like equations.
The red solid line is the plot of the truncated expansion
\eqref{eq:<N>_exp} up to $n=9$ where we have used the exact values of
$\Tr \rho^n$.
The green dashed line represents the perturbative result
\eqref{eq:<N>_pert}.
We find that our numerical results are close to the perturbative
result \eqref{eq:<N>_pert} in the large $\mu$ limit while in the limit
$\mu \to -\infty$ the expansion \eqref{eq:<N>_exp} agrees with the
numerical results as expected.
The TBA  interpolates both results smoothly.
This is not surprising because the TBA results contain all the
non-perturbative corrections, which are neglected in
\eqref{eq:<N>_pert}.

Finally, let us give a comment on an underlying technical issue in
this TBA system.
As was mentioned in \cite{Z}, the iterative solution of the TBA
equations describing the Fredholm determinant converges only when $z$
is in the regime $|z|<1/\lambda_0$ where $\lambda_0$ is the largest
eigenvalue of the kernel $\rho$.
If $|z|$ crosses over this value, the iteration method does not work
any more.
In our interested case $k=1$, the bound of the convergence exists at
$|z|=1/\lambda_0 \approx 7.21$ ($\mu\approx 1.97$).

\section{Discussion}
In this paper, we have studied the Fermi gas quantum mechanics
associated to the ABJM matrix model.
We have computed $\Tr\rho^n$, which is the ingredient of the grand
partition function, and obtained the partition function $Z(N)$ up to
$N=9$ as a result.
In reducing the number of integrations by commuting coordinates and
momenta, we have found an exact relation concerning the grand
partition function, which is interesting by itself and very helpful
for determining the partition function.
We have also performed various numerical analysis.
All of these results are consistent with our exact results and also
with the perturbative analysis in \cite{MPfree}.

Interestingly,  all the partition functions we obtained are written in
terms of polynomials of $\pi^{-1}$ with rational coefficients.
Similarly, the functions $\phi_\pm^k(q)$ which lead to the
determination of $\Tr\rho^n$ are also expressed in terms of
polynomials of $\pi^{-1}$, $q$, $s_q$, $c_q$, $s_{q/2}$ and $c_{q/2}$
with rational coefficients.
Since we have computed the partition function exactly and integer
numbers (or rational numbers) usually counts some information of the
moduli space, we would like to know what kind of the information we
are counting right now.

We would like to understand better the origin of the anomaly in
commuting the coordinate $q$ and the momentum $p$.
From the analogy of string field theory, typically these kinds of
anomaly appear because of the singularity of the moduli space we are
considering and give some information to the physical problem.
For example, the associativity anomaly \cite{HS} in cubic string field
theory \cite{W} comes from the singularity at the boundary of the
moduli space and relates to the translational invariance in string
field theory.
Also, the twist anomaly \cite{HM} in vacuum string field theory
\cite{RSZ} originates from the divergence at the midpoint interaction
point and contains the information of the tachyon mass and the brane
tension.

We have shown that finding the energy spectrum of our Fermi gas system
boils down to the diagonalization of the Hankel matrix $M$ defined by
\eqref{eq:Hankel_k=1}.
As in the case of the diagonalization of Neumann matrices in open
string field theory \cite{Rastelli:2001hh}, the first step would be to
find a (hidden) symmetry $K$ commuting with $M$: $[M,K]=0$.
It would be very interesting to find the exact spectrum of $M$ and
write down the exact grand partition function.

We showed in \eqref{eq:fac_Xi} that the grand partition function of
the ABJM theory is factorized into two parts.
A similar structure has appeared in the study of holographic dual of
BPS black holes \cite{DGOV}, and it was argued that going to the fixed
charge sector leads to the entanglement of two sectors. 
Such entangled state has an interpretation of branched tree of baby
universes and the Catalan number appears as the number of branchings
\cite{DGOV}.
It would be interesting to see if the factorized grand partition
function of the ABJM theory has some interpretation on the bulk
gravity side.

We also would like to remark  on an analogy with the so-called ODE/IM
correspondence \cite{DT,BLZ2}.
In \cite{Vo1,Vo2,Vo3}, it was found that spectral determinants for
certain Schr\"odinger equations satisfy functional relations.
In \cite{DT,BLZ2}, these spectral determinants were identified with
Baxter's $Q$-functions of the corresponding integrable system, and the
functional relations of the spectral determinants just correspond to
the quantum Wronskian relations \cite{BLZ1}.
Our analysis implies that there is a non-trivial relation between the
two determinants $\Xi_\pm(z)$ because the total determinant $\Xi(z)$
is fixed by the information of either of the two sectors $\rho_\pm$.
It would be interesting to explore whether $\Xi_\pm(z)$ satisfy a
functional relation like the quantum Wronskian relation or not.

Recently, the expectation values of some BPS Wilson loops of the ABJM
theory were  analyzed using the Fermi gas formalism \cite{KMSS}.
It was found that the semiclassical expansion of such Wilson loop
expectation values is divergent for $k=1,2$.
We hope that our formalism will give some improvements for the
computation of Wilson loops.

\vskip7mm
\centerline{\bf Acknowledgements}
\noindent
We are grateful to Hiroyuki Hata, Masazumi Honda, Kazuo Hosomichi,
Hiroaki Kanno, Tomoki Nosaka, Satoru Odake, Ryu Sasaki, Naoki Sasakura
for useful discussions.
Y.H. would like to thank Junji Suzuki for telling him the Mathematica
code to solve TBA equations by using FFT algorithm.
Y.H. is grateful to the hospitality of the IPhT, CEA/Saclay and S.M. is
grateful to the hospitality of YITP at Kyoto University.
The work of Y.H. is supported in part by the JSPS Research Fellowship
for Young Scientists, while the work of K.O. is supported in part by
JSPS Grant-in-Aid for Young Scientists (B) \#23740178.

\appendix
\section{Recurrence relations for $\phi_\pm^k(q)$}\label{app:rec-phi}
Let us derive the recurrence relations \eqref{eq:phipm_rec}.
Using \eqref{eq:phipm-def}, we find
\begin{align}
\phi_\pm^k(q)&=\frac{1}{E_\pm(q)}\int \! \frac{dq'}{2\pi}\frac{dq''}{2\pi}
\rho_\pm(q,q')\rho_\pm^{k-1}(q',q'')E_\pm(q'') \nonumber\\
&=\frac{1}{E_\pm(q)}\int \! \frac{dq'}{2\pi}
\rho_\pm(q,q')E_\pm(q')\phi_\pm^{k-1}(q').
\end{align}
Using the identity
\begin{align}
\frac{1}{\cosh q+\cosh q'}
=\frac{1}{2\cosh\frac{q}{2}\cosh\frac{q'}{2}}(c_{q-q'}+c_{q+q'}),
\end{align}
we find that the kernel is written as
\begin{align}
E_+(q)^{-1}\rho_+(q,q')E_+(q')
&=\frac{\cosh\frac{q'}{2}}{2(\cosh q+\cosh q')}
=c_q\frac{c_{q-q'}+c_{q-q'}}{2},
\nonumber\\
E_-(q)^{-1}\rho_-(q,q')E_-(q')
&=\frac{\sinh^2\frac{q'}{2}}{2\cosh\frac{q'}{2}(\cosh q+\cosh q')}
=c_q\frac{c_{q-q'}+c_{q+q'}}{2}t_{q'}^2.
\end{align}
Since $\phi_\pm^{k}(q)$'s are even functions of $q$, we get
\begin{align}
\phi_+^k(q)
&=\int\!\frac{dq'}{2\pi}c_q\frac{c_{q-q'}+c_{q+q'}}{2}\phi_+^{k-1}(q')
=c_q\int\!\frac{dq'}{2\pi}c_{q-q'}\phi_+^{k-1}(q'),\nonumber\\
\phi_-^k(q)
&=\int\!\frac{dq'}{2\pi}c_q\frac{c_{q-q'}+c_{q+q'}}{2}t_{q'}^2\phi_-^{k-1}(q')
=c_q\int\!\frac{dq'}{2\pi}c_{q-q'}t_{q'}^2\phi_-^{k-1}(q'),
\end{align}
which are the recurrence relations \eqref{eq:phipm_rec}.

\section{Hankel representations of $\phi_+^k(q)$ and
$\tilde{\phi}_+^k(q)$}
\label{app:phi-Hankel}
Let us derive \eqref{eq:phi-Hankel}.
We expand $\phi_+^k(q)$ and $\tilde{\phi}_+^k(q)$ in $t_q=\tanh(q/2)$
\begin{align}
\phi_+^k(q)=4c_q^2 \sum_{n=0}^\infty t_q^{2n} \alpha_n^{(k)},\quad
\tilde{\phi}_+^k(q)=\frac{c_q}{2} \sum_{n=0}^\infty t_q^{2n} \beta_n^{(k)},
\label{eq:phi_exp_unknown}
\end{align}
We would like to show
\begin{align}
\alpha_n^{(k)}=\sum_{m=0}^\infty (M_+^k)_{nm},\quad
\beta_n^{(k)}=(M_+^{k-1})_{n0}.
\label{eq:alpha&beta}
\end{align}
Since $\phi_+^0(q)=1$ and $\tilde{\phi}_+^1(q)=c_q/2$, one immediately
finds
\begin{align}
\alpha_n^{(0)}=1,\quad \beta_n^{(1)}=\delta_{n0}.
\label{eq:alpha&beta_ini}
\end{align}
Let us substitute \eqref{eq:phi_exp_unknown} into the recurrence
relation \eqref{eq:phi_rec}.
Using the expansion
\begin{align}
c_{q-q'}=2 c_q c_{q'} \sum_{n=0}^\infty t_q^n t_{q'}^n,
\end{align}
we obtain
\begin{align}
\phi_+^{k}(q)
=c_q \int\frac{dq'}{2\pi}
\( 2 c_q c_{q'} \sum_{n=0}^\infty t_q^n t_{q'}^n \)
\cdot 4c_{q'}^2 \sum_{\ell=0}^\infty t_{q'}^{2\ell} \alpha_\ell^{(k-1)}
\end{align}
The integral over $q'$ causes the Hankel matrix because of
\eqref{eq:Hankel1}.
Thus we find
\begin{align}
\phi_+^{k}(q)
=4c_q^2 \sum_{\ell, n=0}^\infty t_q^{2n}(M_+)_{n \ell}\alpha_\ell^{(k-1)}.
\end{align}
Comparing this equation with \eqref{eq:phi_exp_unknown}, we obtain
\begin{align}
\alpha_n^{(k)}=\sum_{\ell=0}^\infty (M_+)_{n\ell}\alpha_{\ell}^{(k-1)}.
\end{align}
From this relation and the initial condition
\eqref{eq:alpha&beta_ini}, one can show that $\alpha_n^{(k)}$ is given
by \eqref{eq:alpha&beta}.
The derivation of $\beta_n^{(k)}$ is straightforward.

\section{Explicit forms of $\rho_\pm^n(q_1,q_2)$}\label{app:explicit_rhopm}
Let us summarize the explicit forms of $\rho_\pm^n(q_1,q_2)$ up to
$n=4$:
\begin{align}
\rho_+^2(q_1,q_2)&=\frac{1}{16}
\frac{1}{\sqrt{\cosh(\frac{q_1}{2})\cosh(\frac{q_2}{2})}}
\frac{1}{\cosh(\frac{q_1}{2})+\cosh(\frac{q_2}{2})},\\
\rho_+^3(q_1,q_2)&=\frac{1}{16\pi}
\frac{\sqrt{\cosh(\frac{q_1}{2})\cosh(\frac{q_2}{2})}}
{\cosh q_1+\cosh q_2}
\(\frac{q_1}{\sinh q_1}+\frac{q_2}{\sinh q_2}
-\frac{\pi}{2\cosh(\frac{q_1}{2})\cosh(\frac{q_2}{2})}\),\\
\rho_+^4(q_1,q_2)&=\frac{1}{128\pi}
\frac{1}{\sqrt{\cosh(\frac{q_1}{2})\cosh(\frac{q_2}{2})}}
\frac{1}{\cosh(\frac{q_1}{2})-\cosh(\frac{q_2}{2})}
\(\frac{q_1}{\sinh q_1}-\frac{q_2}{\sinh q_2}\)\notag\\
&\quad
+\frac{1}{256}
\frac{1}{\cosh^{3/2}(\frac{q_1}{2})\cosh^{3/2}(\frac{q_2}{2})},
\end{align}
and
\begin{align}
\rho_-^2(q_1,q_2)&=\frac{1}{16}
\frac{\tanh(\frac{q_1}{4})\tanh(\frac{q_2}{4})}
{\sqrt{\cosh(\frac{q_1}{2})\cosh(\frac{q_2}{2})}
(\cosh(\frac{q_1}{2})+\cosh(\frac{q_2}{2}))},\\
\rho_-^3(q_1,q_2)&=\frac{1}{16\pi}
\frac{\sinh(\frac{q_1}{2})\sinh(\frac{q_2}{2})}
{\sqrt{\cosh(\frac{q_1}{2})\cosh(\frac{q_2}{2})}(\cosh q_1+\cosh q_2)}\notag\\
&\qquad\qquad\times
\(\frac{q_1}{\sinh q_1}+\frac{q_2}{\sinh q_2}
-\frac{\pi}{2\cosh(\frac{q_1}{2})\cosh(\frac{q_2}{2})}\)\notag\\
&\quad-\frac{1}{64}\frac{\tanh(\frac{q_1}{4})\tanh(\frac{q_2}{4})}
{\cosh^{3/2}(\frac{q_1}{2})\cosh^{3/2}(\frac{q_2}{2})}, \\
\rho_-^4(q_1,q_2)&=\frac{1}{256\pi}
\frac{\tanh(\frac{q_1}{4})\tanh(\frac{q_2}{4})}
{\sqrt{\cosh(\frac{q_1}{2})\cosh(\frac{q_2}{2})}}
\(\frac{1}{\sinh^2(\frac{q_1}{4})\sinh^2(\frac{q_2}{4})}-
\frac{\pi}{\cosh(\frac{q_1}{2})\cosh(\frac{q_2}{2})}\) \notag \\
&\quad
+\frac{1}{512\pi}\frac{\tanh(\frac{q_1}{4})\tanh(\frac{q_2}{4})}
{\sqrt{\cosh(\frac{q_1}{2})\cosh(\frac{q_2}{2})}
(\cosh(\frac{q_1}{2})-\cosh(\frac{q_2}{2}))}\notag \\
&\qquad\qquad\times
\(\frac{q_1 \coth(\frac{q_1}{4})}
{\cosh(\frac{q_1}{2})\sinh^2(\frac{q_1}{4})}
-\frac{q_2 \coth(\frac{q_2}{4})}
{\cosh(\frac{q_2}{2})\sinh^2(\frac{q_2}{4})}\).
\end{align}
These matrix elements are directly derived from \eqref{eq:phi_+},
\eqref{eq:phi_-} by using \eqref{eq:rho_odd} and
\eqref{eq:rho_even}.\footnote{
Similarly we also obtain $\rho_\pm^n(q_1,q_2)$ for $n=5,6,7$.
Since it can be obtained without difficulty, we shall omit their
expressions here due to the complexity.}
Using these matrix elements, we can also compute $\Tr \rho_\pm^{2n+1}$
up to $n=3$ in addition to $\Tr \rho_\pm^{2n}$ as given in
\eqref{eq:Trrhop} and \eqref{eq:Trrhom}.
The results are given by
\begin{align}
\Tr \rho_+&=\frac{\sqrt{2}}{8},\quad
\Tr \rho_+^3=\frac{3-2\sqrt{2}}{64},\quad
\Tr \rho_+^5=\frac{(5-8\sqrt{2})\pi+20}{2^{10}\pi},\nonumber\\
\Tr \rho_+^7&=\frac{(16\sqrt{2}-21)\pi^2-14\pi+28}{2^{13}\pi^2},
\label{eq:Trrhop_odd}
\end{align}
and
\begin{align}
\Tr \rho_-&=\frac{2-\sqrt{2}}{8},\quad
\Tr \rho_-^3=\frac{(1+2\sqrt{2})\pi-12}{64\pi},\quad
\Tr \rho_-^5=\frac{(8\sqrt{2}-9)\pi^2-20\pi+40}{2^{10}\pi^2},\nonumber\\
\Tr \rho_-^7&=\frac{-9(16\sqrt{2}+3)\pi^3+518\pi^2+756\pi-336}
{2^{13}\cdot3^2\pi^3}.
\label{eq:Trrhom_odd}
\end{align}
Interestingly, in the above results the terms with $\sqrt{2}$
appear.
Such terms are canceled out in $\Tr \rho^{2n+1}$, and 
$\Tr \rho^{2n+1}$ do not contain them.

\end{document}